\documentclass[aps,twocolumn,prb,showpacs,preprintnumbers,superscriptaddress,amsmath,amssymb]{revtex4}
\usepackage{graphicx}
\usepackage{dcolumn}
\usepackage{color}
\usepackage[usenames,dvipsnames]{xcolor}
\usepackage{bm}
\numberwithin{equation}{section}

\newcommand{\dpsi}{\psi^\dagger}

\newcommand{\tphi}{\tilde{\phi}}

\newcommand{\hh}{\hat{H}}

\newcommand{\ud}{\,\mathrm{d}}
\newcommand{\bg}{\bar{g}}
\newcommand{\tg}{\bar{\gamma}}
\newcommand{\im}{\mathrm{i}}
\newcommand{\e}{\textrm{e}}

\DeclareMathOperator{\hside}{\Theta}

\DeclareMathOperator{\sgn}{sgn}
\DeclareMathOperator{\Order}{O}

\begin{document}

\title{Conducting fixed points for inhomogeneous quantum wires:\\ a conformally invariant boundary theory}

\author{N.~Sedlmayr}
\email{nicholas.sedlmayr@cea.fr}
\author{D.~Morath}
\author{J.~Sirker}
\author{S.~Eggert}
\affiliation{Department of Physics and Research Center OPTIMAS, University of Kaiserslautern,
D-67663 Kaiserslautern, Germany}
\author{I.~Affleck}
\affiliation{Department of Physics and Astronomy, The University of British Columbia,
Vancouver, BC V6T 1Z1, Canada}

\begin{abstract}
  Inhomogeneities and junctions in wires are natural sources of
  scattering, and hence resistance. A conducting fixed point usually
  requires an adiabatically smooth system.  One notable exception is
  ``healing'', which has been predicted in systems with special
  symmetries, where the system is driven to the homogeneous fixed
  point.  Here we present theoretical results for a different type of conducting fixed
  point which occurs in inhomogeneous wires with an abrupt jump in
  hopping and interaction strength.  We show that it is always
  possible to tune the system to an unstable conducting fixed point
  which does not correspond to translational invariance.  We
  analyze the temperature scaling of correlation functions at and near
  this fixed point and show that two distinct boundary exponents
  appear, which correspond to different effective Luttinger liquid
  parameters.  Even though the system consists of two separate
  interacting parts, the fixed point is described by a single
  conformally invariant boundary theory. We present details of the
  general effective bosonic field theory including the mode expansion
  and the finite size spectrum. The results are confirmed by numerical
  quantum Monte Carlo simulations on spinless fermions.  We predict  characteristic
experimental signatures of the local density of states near junctions.
\end{abstract}


\maketitle

\section{Introduction}\label{sec_intro}
Transport in quantum wires is a rich field bringing together
conductivity
experiments \cite{Liang2001,Javey2003,Yacoby1996,Steinberg2008,Tarucha1995}
and Luttinger liquid theory which describes the crucial
electron-electron interaction effects in one
dimension.\cite{Tomonaga1950,Luttinger1963,Giamarchi2004} Scattering
from a single impurity or other inhomogeneities, for example, becomes
renormalized by the interaction and can lead to insulating behavior
at low temperatures even for weak
impurities.\cite{Kane1992,Kane1992a,Eggert1992,Furusaki1993,Pereira2004a,Sedlmayr2011b}

In order to determine the conductivity of a one-dimensional wire it is
necessary to couple it to some leads or reservoirs, normally a two
dimensional electron gas (2DEG). Such a set up can be most readily
described as an inhomogeneous wire, in which the 2DEGs are modeled as
non-interacting wires. In this case the conductance is usually
controlled by the parameters of the lead rather than of the
wire,\cite{Yue1994,Safi1995,Maslov1995,Ogata1994,Wong1994,Chamon1997,Imura2002,Enss,Rech2008,Rech2008a,Gutman2010a,Thomale2011,Sedlmayr2012a,Sedlmayr2013}
in contrast to what a naive calculation on an {\it infinite} interacting wire
would suggest.   The conductance for perfect adiabatic contacts and wires
can be understood by the decomposition of an electron into fractional
charges.\cite{Safi1995,Safi1999} Additional relaxation processes which
take place within the interacting region of the wire do, however, lead
to a resistance which is affected by the wire parameters. The
resistance due to impurity scattering \cite{Furusaki1996} or phonon
scattering \cite{Sedlmayr2013} within the interacting wire, for
example, will in general depend both on the Luttinger liquid parameter
of the leads and the wire.

In this paper we consider
the intrinsic scattering from the junctions between
the wire and leads, which is generically present due to the abrupt change of parameters
even for otherwise perfect ballistic connections.
This scattering is renormalized by the
interaction,\cite{Furusaki1996} leading to a vanishing dc conductance
in the low temperature limit for repulsive interactions within the
wire. However, perfect conductance is still possible by tuning the
parameters on the two sides of the junctions as has been
analyzed in detail for a particle-hole symmetric
model.\cite{Sedlmayr2012a} In this case a line of conducting fixed
points in parameter space exists as only one relevant backscattering
operator is permitted by symmetry which can always be tuned to zero.
Here we generalize to the more experimentally relevant case where
particle-hole symmetry is no longer present. Even in this more general
case we still find a line of conducting fixed points provided the underlying
microscopic theory has certain local symmetry properties.
Even though the systems under consideration are inhomogeneous,
it is possible to characterize the
fixed points by a single conformally invariant boundary theory with a characteristic
 mode expansion
 and finite size spectrum. The results are confirmed by numerical
 quantum Monte Carlo (QMC) simulations on spinless fermions.  Characteristic
experimental signatures for the local density of states near junctions can be predicted.

For conductivity experiments we must typically consider a system with two junctions, one at
each end of an interacting wire where it is connected to the leads (e.g.~2DEGs).
These junctions are
intrinsic sources of inhomogeneity, but in most cases the junctions do not
influence each other since the length of the wire is much
larger than the coherence length $u\beta$, where $\beta$ is the
inverse temperature and $u$ the velocity of the collective
excitations.   For our purpose
to make predictions for the backscattering and the local behavior
near the leads, it is therefore sufficient to analyze one junction between a
lead and a wire.

As an introduction in Sec.~\ref{sec_nonint} we consider an idealized
junction in a non-interacting lattice model and discuss the
applicability of a narrow band approximation. In
Sec.~\ref{sec_interacting} we start from a microscopic interacting
model and demonstrate how the backscattering terms arise, and then
introduce the general effective bosonic field theory. Focusing on
abrupt junctions connecting otherwise homogeneous wires,
we examine
the renormalization group flow of perturbing operators in the model.
We discuss the locations of the unstable conducting fixed points in
relation to the symmetry properties of the underlying microscopic
model. Finally in Sec.~\ref{sec_cft} we describe the conformally
invariant boundary theory for the conducting fixed point and the
scaling of the local correlation functions at the boundary. In
Sec.~\ref{conclusions} we conclude.

\section{Non--interacting models}\label{sec_nonint}
Before considering the interacting model it is instructive to analyze
the backscattering seen in inhomogeneous systems of free particles,
where exact results are obtainable and can be compared directly with
low energy approximations. We start with a lattice model of
non-interacting spinless fermions described by the Hamiltonian
\begin{equation}\label{freeh}
\hh_0 - \mu N =-\sum_j[t_j(\psi^\dagger_j\psi_{j+1}+\textrm{H.c.})-V_j\psi^\dagger_j\psi_j]\,.
\end{equation}
$\psi^\dagger_j$ creates a particle at site $j$, $t_j$ and $V_j$ are
the position dependent hopping elements and local potential energy
respectively, and $N$ the total particle number. 
 We set $\hbar=1$ and include in the following the chemical
potential $\mu$ in the local potential energy $V_j$. Generically we
consider situations in which we have two homogeneous regions on the
left ($j\leq j_\ell$) and right ($j\geq j_r$) side of the wire.  In
these asymptotic regions the plane-wave solutions have the same energy
so the parameters are related by
\begin{equation}
\label{energy_match}
-2t_\ell\cos[k_\ell a]+V_\ell =-2t_r\cos[k_ra]+V_r\,,
\end{equation}
with $k_{\ell,r}$ the momenta, $V_{\ell,r}$ the potential, and
$t_{\ell,r}$ the hopping on the left ($\ell$) and right ($r$) side. We have also introduced
the lattice spacing $a$. We consider a wave-function incident from the left
\begin{equation}\label{scatter}
\psi_j=\left\{\begin{array}{ll}
		  \e^{\im k_\ell j}+R\e^{-\im k_\ell j}\,,& j\leq j_\ell \\
		  Te^{\im k_rj}\,,& j\geq j_r
                \end{array}\right. .
\end{equation}
The region from $j_\ell$ to $j_r$ is the region of inhomogeneity describing the junction.

There are two velocities
\begin{equation}
u_i\equiv 2at_i\sin[k_ia]\,,
\end{equation}
$i=\{\ell,r\}$, and current conservation implies
\begin{equation}\label{cc}
(1-|R|^2)u_\ell =|T|^2u_r.
\end{equation}
It is natural to refer to $R=0$ as ``perfect transmission'', although
this does not  necessarily maximize $|T|^2$. A reasonable definition of
perfect transmission would be maximizing the outgoing current on the right
for a given value of the incoming current from the left, $u_\ell$; that is, maximizing
$|T|^2u_r/u_\ell$.  Noting that $|T|^2u_r/u_\ell =1-|R|^2$ we see that the condition for perfect transmission equivalently corresponds
to minimizing $|R|^2$. This can also be seen by considering the Landauer transmission, see Appendix \ref{App_Landauer}.

In general, accurate results cannot be obtained by ignoring states
far from the Fermi energy. This can be seen from the fact that the
off-diagonal components of the $T$-matrix, $T_{k,k'}$ are
non-negligible when $|k'|$ is not close to $|k|$.  This implies a
non-negligible mixing of low energy states with high energy ones due
to scattering near the interface. However, in certain limits, a narrow
band theory can be used, in which we keep only a narrow band of
states, of width $\Lambda \ll k_F$, where $k_F$ is the Fermi momentum,
and linearize the dispersion relation.  This can be justified in one
of two cases. a) If all potential energy terms $V_i$ and all hopping terms $t_i$ are nearly equal,
including the asymptotic ones $t_\ell\approx t_r$.  This corresponds to the adiabatic limit where a local density approximation suffices. b) If there are
one or more very weak hopping terms separating otherwise uniform
chains. In this latter case the ratio $t_\ell/t_r$ can be arbitrary.
These are the limits of weak backscattering or weak tunneling.
Starting with the unperturbed basis of translationally invariant wave-functions, or wave-functions
vanishing at the interface respectively, a small perturbation only mixes states with
energy differences of order of magnitude of the perturbation.

In these cases we may keep only a narrow band of states near zero
energy and introduce left and right moving fields in the usual way,
\begin{equation}\label{narrow}
\frac{\psi_j}{\sqrt{a}}\approx \e^{\im k_{F,x}x}\psi_+(x)+\e^{-\im k_{F,x}x}\psi_-(x)\,,
\end{equation}
with $x=aj$ a continuous variable and $k_{Fx}$ being the Fermi
momentum in the left, $k_{F,x\leq aj_\ell}=k_{F\ell}$, or right,
$k_{F,x\geq aj_r}=k_{Fr}$, of the wire.

Here we want to consider only the simplest model for a junction while
various other types of junctions are discussed in Appendix
\ref{App_NonInt}. In the simplest model two homogeneous regions are
connected at one site such that
\begin{eqnarray}\label{abrupt}
t_j&=&\left\{\begin{array}{ll}t_\ell\,,&  j<0\\
              t_r,&  j\geq 0
             \end{array}\right.\\\nonumber
V_i&=&\left\{\begin{array}{ll}V_\ell \,,&  j<0\\
              V_r,& j> 0
             \end{array}\right.
\end{eqnarray}
and $V_0$ is kept as a free parameter.
The reflection amplitude is determined by the Schr\"odinger equation for the central site and results in
\begin{equation}\label{sq}
R=-{a(V_\ell+V_r-2V_0)-i(u_\ell-u_r)\over a(V_\ell+V_r-2V_0)+i(u_\ell+u_r)}
\end{equation}
The conditions for perfect transmission are therefore
\begin{eqnarray}
u_\ell &=&u_r\nonumber\,,\textrm{ and}\\
V_0&=&(V_\ell +V_r)/2\,.\label{perf}
\end{eqnarray}
When these conditions are satisfied, $R=0$ and $|T|^2=1$. Curiously,
the maximum possible value of $|T|^2$ actually occurs when
$V_0=(V_\ell +V_r)/2$ and $u_r=0$, in which case $|R|=1$ and
$|T|=2$. But in this case the current is actually zero on both sides,
so calling this perfect transmission would seem inappropriate. The
existence of the two conditions \eqref{perf} for perfect conductance
is related to the breaking of particle--hole symmetry, see Sec.~\ref{sec_abrupt}.

Next, we consider the abrupt junction of Eq.~(\ref{abrupt}) in a
narrow band approximation setting $t_\ell=t-\delta t$ and
$t_r=t+\delta t$, with $|\delta t|\ll t$. When $\delta t=0$ we obtain
the usual free, translationally invariant Dirac fermion model, with
uniform velocity $u_0=2at\sin k_F$. Here we treat the $\delta t$ term as a
perturbation. Using the separation into right and left moving
fields, Eq.~\eqref{narrow}, the backscattering at the junction is given
by
\begin{eqnarray}
&& \!\!\!\!\!\!\!\!\!\delta \hh \approx -a\left[ \sum_{j=-\infty}^{-1} (2t_\ell e^{i k_{F\ell}a} -V_\ell) e^{2ik_{F\ell}ja} + 2t_r e^{i k_{Fr}a} \right.\nonumber \\
&& \!\!\!\!\!\!\!\!\!\!\! \left . -V_0 + \sum_{j=1}^{\infty} (2t_r e^{i k_{Fr}a} -V_r) e^{2ik_{Fr}ja} \right] \psi_-^\dagger \psi_+ +\textrm{H.c.} 
\end{eqnarray}
Since $\psi_-(x)$ and $\psi_+(x)$ are assumed to vary slowly on the scale of $k_{F\ell/r}^{-1}$ the
oscillating terms in the bulk cancel, leaving only the contributions at $x=0$.
We may then write the local backscattering at $x=0$ as
\begin{equation}
\delta \hh\approx 2\pi\im\lambda \psi^\dagger_-\psi_+(x=0)+\textrm{H.c.}
\end{equation}
with (see also Appendix E)
\begin{eqnarray}
\label{lambda_half}
{\rm Re}\,\lambda &=& \frac{a}{2\pi}\left(\frac{t_\ell}{\sin[k_{F\ell} a]} - \frac{t_r}{\sin[k_{Fr} a]}\right) \\
&-& \frac{a}{4\pi} \left(V_\ell\cot[k_{F\ell}a] -V_r\cot[k_{Fr} a]\right) \nonumber \\
&=& \frac{a}{2\pi}\left( t_\ell \sin[k_{F\ell} a] -t_r\sin[k_{Fr}a]\right), \nonumber \\
{\rm Im}\,\lambda &=& \frac{a}{4\pi}\left( V_\ell+V_r-2V_0\right) , \nonumber
\end{eqnarray}
where we have used Eq.~\eqref{energy_match} to simplify the real part.
We see that the scattering amplitude $\lambda$ is real if the local
potential energies are equal, $V_0=V_\ell = V_r$. This is surprising
because for any non-zero local potential the problem is no longer
particle-hole symmetric. In App.~\ref{App_NonInt} we show that this is
a special property of the junction \eqref{abrupt} and does not hold in
general.  Finally, we can use the fact that we are treating the
difference in hopping $\delta t$ perturbatively and approximate
$k_{F\ell}\approx k_{Fr}\approx k_F$ in which case the real part of
the scattering amplitude further simplifies,
\begin{equation}
{\rm Re}\, \lambda = - \frac{a\,\delta t}{\pi}\sin[k_F a] = \frac{u_\ell -u_r}{4\pi} \, ,
\end{equation}
where the difference in velocities on the two sides of the junction is
given by $u_r-u_\ell =4a\delta t\sin[k_Fa]$. We see that 
for $V_0\approx V_\ell\approx V_r$ and $u_\ell\approx u_r$, required for the
narrow band approximation to be valid, the result for the scattering
amplitude $\lambda$ is fully consistent with the exact result for the
reflection amplitude \eqref{sq} by using the general relation
$R=4\pi\lambda/(u_r+u_\ell)$ between these two quantities in this
limit. In Sec.~\ref{sec_abrupt} we will discuss how the narrow band
calculation for this type of junction can be extended to the
interacting case using bosonization.

\section{Interacting model}\label{sec_interacting}
As a microscopic interacting model we use the Hamiltonian
$\hh=\hh_0+\hh_I$, where $\hh_0$ is given by Eq.~\eqref{freeh} and
\begin{equation}\label{spin_int}
\hh_I=\sum_{j} U_j:\dpsi_{j}\psi_{j}::\dpsi_{j+1}\psi_{j+1}\,:
\end{equation}
for interactions with a position dependent nearest neighbor
interaction strength $U_j$. Normal ordered operators are given by
$:\dpsi_j\psi_j:=\dpsi_j\psi_j-\langle0|\dpsi_j\psi_j|0\rangle$, with
$|0\rangle$ the ground state.  It is assumed that the spatial
variation of $U_j$, $t_j$, and $V_j$ in $\hh_0$, is consistent
with the narrow band approximation explained in the preceding section.
Later we will focus on the limiting case of an abrupt jump in the
interaction and hopping parameters at the junction, as used
elsewhere.\cite{Safi1995,Maslov1995,Furusaki1996,Safi1999,Sedlmayr2012a}

In order to find the underlying low energy bosonic theory, we first
need to linearize the spectrum. Analogously to the normal Luttinger
liquid theory,\cite{Tomonaga1950,Luttinger1963,Giamarchi2004} one can
linearize around the bulk band structure in the left and right regions of the wire.\cite{Sedlmayr2012a} Linearization is performed around the
Fermi momenta $k_{F,x}$ for left and right movers:
\begin{equation}\label{lin}
\frac{\psi_j}{\sqrt{a}}=\psi(x)=\sum_{\alpha=\pm}\e^{\im \alpha k_{F,x}x}\psi_\alpha(x)\,,
\end{equation}
with the appropriate commutation relations $\left[\psi_\alpha(x),\psi_\beta(x')\right]_+=0$ and $\left[\psi_\alpha(x),\dpsi_\beta(x')\right]_+=\delta_{\alpha\beta}\delta(x-x')$. 
Here $k_{F,x}$ is defined by $-2t\cos k_{F,x}+V_x=0$.  
Note that it is not necessary to assume that $k_{F\ell}\approx k_{Fr}$.

After linearization of the free Hamiltonian we find
\begin{eqnarray}
\hh_0&=&-\int\ud x\sum_{\alpha=\pm}at_x\big[\e^{\im \alpha \kappa^-_x}\dpsi_{\alpha}(x)\partial_x\psi_\alpha(x)+\textrm{H.c.}\big]\nonumber\\&&\nonumber
-\int\ud x\sum_{\alpha=\pm}\left[2t_x\e^{-2\im\alpha \kappa^+_x}-V_x \e^{-2\im\alpha k_{F,x}}\right]\\&&\qquad\times
\dpsi_\alpha(x)\psi_{-\alpha}(x)\,, \label{h0}
\end{eqnarray}
where the Fermi momenta are determined by
\begin{equation}\label{mu}
V_x=2t_x\cos\left[\kappa^-_x\right]\,,
\end{equation}
and we have defined $\kappa^-_x=k_{F,x+a}(x+a)- k_{F,x}x$ and $2\kappa^+_x=k_{F,x+a}(x+a)+k_{F,x}x$.
Similarly, one can write the linearized interaction as
\begin{eqnarray}\label{linear_int_ham}
\hh_I&=&\sum_{\alpha,\beta=\pm}\int\ud xaU_x\bigg(:\dpsi_\alpha\psi_\alpha(x)::\dpsi_\beta\psi_\beta(x+a):\nonumber\\&&\nonumber+
\e^{-\beta 2ik_{F,x+a}(x+a)}:\dpsi_\alpha\psi_\alpha(x)::\dpsi_\beta\psi_{-\beta}(x+a):\\&&\nonumber+
\e^{-\alpha 2ik_{F,x}x}:\dpsi_\alpha\psi_{-\alpha}(x)::\dpsi_\beta\psi_\beta(x+a):\\&&+
\e^{-\alpha 2ik_{F,x}x-\beta 2ik_{F,x+a}(x+a)}\\&&\nonumber\qquad\times:\dpsi_\alpha\psi_{-\alpha}(x)::\dpsi_\beta\psi_{-\beta}(x+a):\bigg)\,,
\end{eqnarray}
keeping for the moment all of the terms. If the interaction acts homogeneously then many of
the terms can be neglected as they are suppressed by the rapidly
oscillating phases. Due to the inhomogeneity in $U_x$ this is no
longer true and all processes could in principle be important. In fact
we find that umklapp scattering is generically irrelevant under renormalization
group (RG) flow, see Appendix \ref{app_rg}, and to lowest order the backscattering only
renormalizes the single particle backscattering already present in the
non-interacting Hamiltonian.

We bosonize using the local vertex operator \cite{Haldane1981,Haldane1981a}
\begin{equation}
\label{bosform}
\psi_{\alpha}(x)=\frac{1}{\sqrt{2\pi a}}e^{\im \alpha\sqrt{4\pi}[\phi_{\alpha}(x)]}\,.
\end{equation}
We use the following convention: $\phi(x)=\phi_+(x)+\phi_-(x)$ and its adjoint $\tphi(x)=\phi_+(x)-\phi_-(x)$ with the conjugate momentum, $\Pi(x)=\partial_x\tphi(x)$. These fields obey
\begin{eqnarray}
[\phi_+(x),\phi_-(y)]&=&-\frac{\im}{4},\\ \nonumber
[\phi_\alpha(x),\phi_\alpha(y)]&=&\frac{\im\alpha}{4}\sgn(y-x),\textrm{ and}\\ \nonumber
[\phi(x),\Pi(y)]&=&i\delta(x-y)\,.
\end{eqnarray}
Some further useful formulas for bosonization are given in
App.~\ref{H_Appendix}.

The full Hamiltonian $\hh=\hh_0+\hh_I$ can be rewritten in the bosonic representation as a quadratic Hamiltonian, a local backscatterer, and umklapp scattering: $\hh=\hh_b+\hh'+\hh_U$, see App.~\ref{H_Appendix} for details. As already mentioned, away from half-filling the umklapp scattering term $\hh_U$ becomes a local perturbation confined to the regions where $U_j$ is varying, and is then irrelevant under RG flow. It is neglected in the following. We find the quadratic term to be
\begin{equation}\label{spinless_ham}
\hh_b=\int\ud x \frac{u_x}{2}\left(\frac{1}{g_x}(\partial_x\phi)^2+g_x(\partial_x\tphi)^2\right)\,.
\end{equation}
To lowest order we can determine the renormalized velocity
\begin{equation}\label{renvel}
u_x\approx 2at_x\sin[\kappa^-_x]\left(1+\frac{U_x}{\pi t_x}\sin[\kappa^-_x]\right)\,,
\end{equation}
and the Luttinger parameter
\begin{equation}
g_x\approx 1-\frac{U_x}{\pi t_x}\sin[\kappa^-_x]\,.
\end{equation}

The local backscattering from all processes in Eqs.~\eqref{h0} and \eqref{linear_int_ham}
can be summarized in one term
\begin{equation}
\hh'=\sum_{\substack{x=ja\\j\in\mathbb{Z}}}\frac{1}{2\pi \im} \e^{-\im \sqrt{4\pi}\phi(x)-2ik_{F,x}x}\left[\frac{e^{-\im \kappa^-_x}u_x}{a\sin[\kappa^-_x]}-V_x\right]
+\textrm{H.c.}.
\end{equation}
We keep the sum over $x=ja$ here discrete in order to avoid ambiguity as to what the alternating terms are in the continuum limit. This also helps the precise calculation of these sums.

\subsection{An abrupt junction}\label{sec_abrupt}
Let us now focus on the simple junction considered already in the
previous section for the non-interacting case where two semi-infinite
wires are joined at $x=0$ with $t_{x< 0}=t_\ell$, $t_{x\geq 0}=t_r$, and
$U_x$ defined equivalently.  The local potential energy is taken to be
uniform, $V_j=V$, except where explicitly said to the contrary.
The Fermi momenta, $k_{F,x}$, can also be written with a similar
structure as $k_{F,x< 0}=k_{F\ell}$ and $k_{F,x\geq 0}=k_{Fr}$.  In
this system backscattering can be rewritten as
\begin{eqnarray}
\hh'&\approx&\lambda\e^{-\im \sqrt{4\pi}\phi(x=0)}+\textrm{H.c.},\nonumber\\
\lambda&=&-i\sum_{x}\frac{1}{2\pi a} \e^{-2\im k_{F,x}x}\left[\frac{\e^{-\im \kappa^-_x}u_x}{\sin[\kappa^-_x]}-V a\right]\,.\qquad\label{full_lambda}
\end{eqnarray}
With the help of appendix \ref{app_sum}, and noting that for an abrupt jump $\kappa^-_x=k_{F,x}a$, we have to lowest order in the interaction
\begin{eqnarray}\label{lambda}
\lambda&\approx&\frac{1}{2\pi}\left[\frac{t_\ell}{\sin[k_{F\ell}a]}+\frac{U_\ell}{\pi}-\frac{t_r}{\sin[k_{Fr}a]}-\frac{U_r}{\pi}\right]\nonumber\\&&-\frac{V}{4\pi}\left[\cot[k_{F\ell}a]-\cot[k_{Fr}a]\right]\,,
\end{eqnarray}
which generalizes the non-interacting result, Eq.~\eqref{lambda_half}.
As $\lambda$ is real we find that there is no
$\sin[\sqrt{4\pi}\phi(0)]$ operator present at the boundary and the
total backscattering is
\begin{equation}\label{total_bs}
\hh'=2\lambda\cos[\sqrt{4\pi}\phi(0)]\,.
\end{equation}
The perhaps surprising absence of the $\sin[\sqrt{4\pi}\phi(0)]$
operator is connected to the local properties of the Hamiltonian in
the vicinity of the boundary, see App.~\ref{App_NonInt}. As such
there remains only one condition to fulfill for the conducting fixed
point: $\lambda=0$ with $\lambda$ real. 

For $V=0$ when there is particle-hole symmetry present, corresponding
to the mapping $\phi\to-\phi$ and $\tphi\to-\tphi$, it is transparent
that $\sin[\sqrt{4\pi}\phi(0)]$ is forbidden. For $V\neq 0$ we find
that $\lambda$ remains real for the specific junction
considered---analytically to first order in the interaction $U$, see
Eq.~\eqref{lambda}, and numerically for all interactions strengths,
see below. We do not have a simple argument why this is the case and
App.~\ref{App_NonInt} shows that this is in fact not a generic feature
of an abrupt junction.

{\subsection{Local density and compressibility}\label{sec_density}

For the system with an abrupt jump in hopping and interaction
strength it is possible to calculate a variety of properties
perturbatively in the boundary operators using the exact Green's function for the Hamiltonian
\eqref{spinless_ham}, see Eq.~\eqref{green} in the Appendix. In
addition to the dc conductance one can also consider local
properties such as the local density and compressibility of the wire.
For abrupt changes in parameters the local density is known to show
characteristic oscillations, the Friedel
oscillations \cite{Friedel}, which give information
about the interacting correlation functions \cite{Egger95,Eggert1995,Soffing09} and
the strength of the backscattering.\cite{Rommer00,Sedlmayr2012a}
\begin{figure}
\includegraphics*[width=0.99\columnwidth]{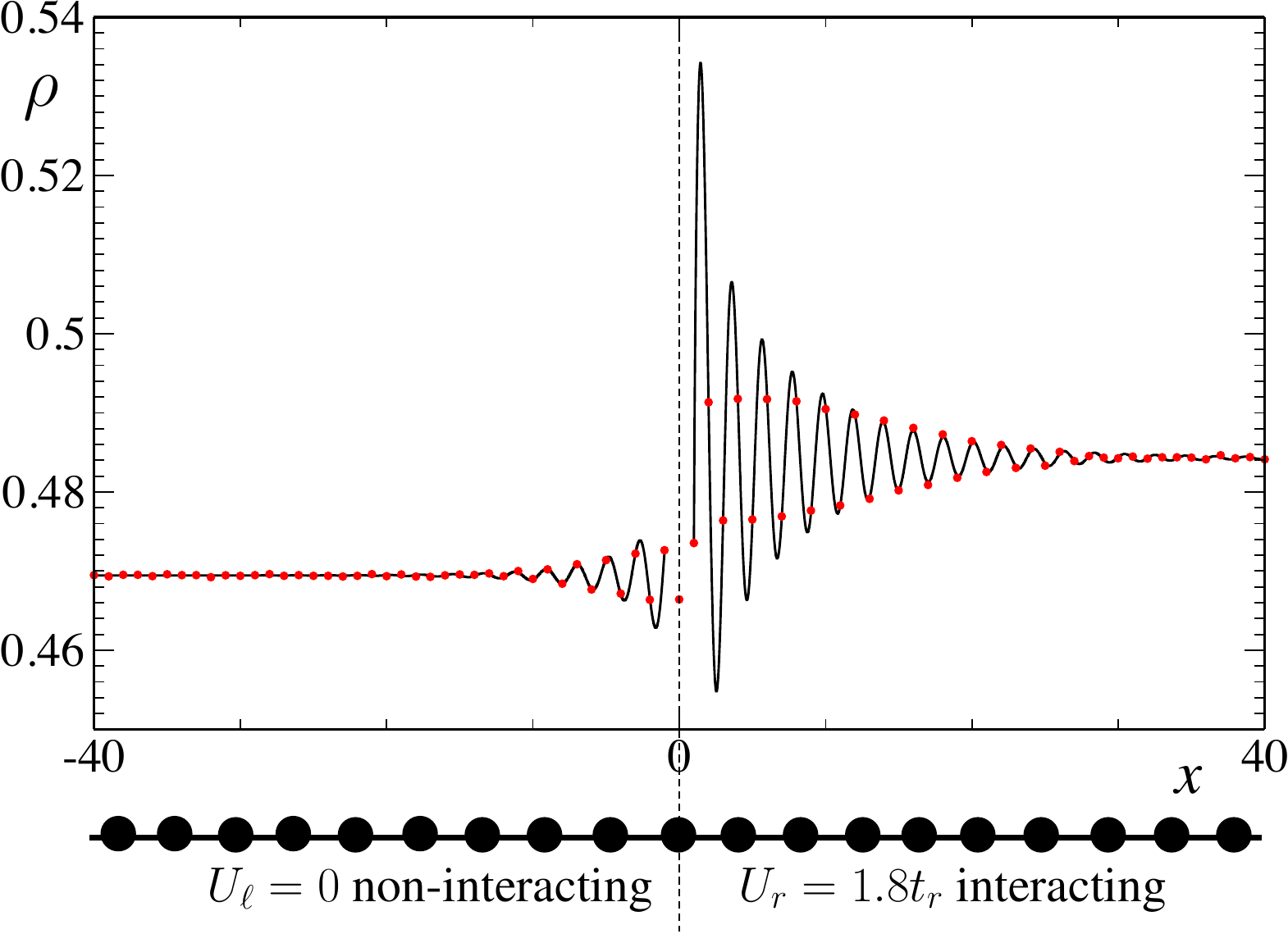}
\caption{(Color online) The full density including Friedel oscillations near the boundary, numerical results (filled circles) are fitted to the analytical result of Eq.~\eqref{dense} (lines) with
  $t_l=1.308t_r$, $U_\ell=0$, and  $U_r=1.8t_r$. The local potential energy is $V=0.25t_r$
and $t_r\beta=10$. Underneath a schematic of the system under consideration is shown.}
\label{density_fig}
\end{figure}

The bosonized density operator for the fermions becomes
\begin{eqnarray}
\label{density_op}
n(x)& =&  n_{0}(x)
-\frac{1}{\sqrt{\pi}}\partial_x\phi(x)\nonumber\\&&+
\frac{\textrm{const.}}{\pi}\sin[2k_{F,x}^*x+\sqrt{4\pi}\phi_x]\,.\quad
\end{eqnarray}
As before we keep the local potential energy constant, $V_x=V$.
The oscillating contribution to the density, \emph{i.e.}~the Friedel oscillations, which are given by
\begin{equation}\label{rhoalt}
\rho_{\rm alt}(x)\equiv\left\langle\frac{\textrm{const.}}{\pi}\sin[2k_{F,x}^*x+\sqrt{4\pi}\phi(x)]\right\rangle\,,
\end{equation}
will be calculated to first order in $\lambda$.
$k_{F,x}^*$ is the renormalized Fermi momentum at finite temperatures which can be found from the bulk density: $\rho_x\equiv\langle n_{0,x}\rangle=k_{F,x}^*/\pi$.

For this we require the following integral
\begin{eqnarray}
\tau(x)&\equiv&
2\int_0^\beta\ud\tau \langle\cos[\sqrt{4\pi}\phi(x,0)]\cos[\sqrt{4\pi}\phi(0,\tau)]\rangle
\nonumber\\&=&
\int_0^\beta\ud\tau \e^{2\pi\left[G(x,0;\tau)-G(0,0;0)\right]}
\\\nonumber&=&\frac{1}{T}\left(\frac{4\pi Ta}{u_x}\right)^{\bg}
\left(\frac{u_x}{2\pi aT} \sinh\left[\frac{2\pi Tx}{u_x}\right]\right)^{-{g_x}}P_{-{\bg}}(z)
\end{eqnarray}
which has been calculated using the Green's function in Appendix
\ref{app_rg}. We introduced
\begin{equation}
z\equiv \coth\left[\frac{2\pi Tx}{u_x}\right]\,,
\end{equation}
and $P_{l}(z)$ is the Legendre function. This gives
\begin{eqnarray}\label{dense}
\rho_{\rm alt}(x)&=&-\frac{\textrm{const.}}{\pi}\int_0^\beta\ud\tau\langle\sin[2k_{F,x}^*x+\sqrt{4\pi}\phi(x)]\hh'\rangle\nonumber\\
&=&-\lambda\frac{\textrm{const.}}{\pi^2 a}\tau(x)\sin[2k_{F,x}^*x]
\end{eqnarray}
In order to test the calculations we have developed a quantum Monte Carlo (QMC) code
using a stochastic series expansion (SSE) with directed
loops.\cite{Syljuaasen2002,Dorneich2001} In Figs.~\ref{density_fig}
and \ref{density_fig2} we show a comparison of this analytical result
with the outcome of QMC simulations on spinless Fermions. Even for a very
large jump in parameters the fit remains very good. Note that what is
seen in the local density and compressibility profiles, see below, is
an interplay between the shape of $\tau(x)$ and the incommensurate
oscillations from $\sin[2k_{F,x}^*x]$. For the fitting procedure
between the analytical and numerical results there are two parameters.
The first is the amplitude of the effect due to the unknown constant
in Eq.~\eqref{rhoalt} and the cutoffs in the field theory.  The second
is a small offset in position, $\rho_{\rm alt}(x-\bar a)$, due to an
effective width of the scattering center, with $\bar a$ being of the
order of the lattice spacing $a$. The Luttinger parameters $g_{\ell,r}$ can be found from Bethe ansatz.\cite{tak99,HubbardBook,PereiraSirkerJSTAT,SirkerLL}
\begin{figure}
\includegraphics*[angle=270,width=0.99\columnwidth]{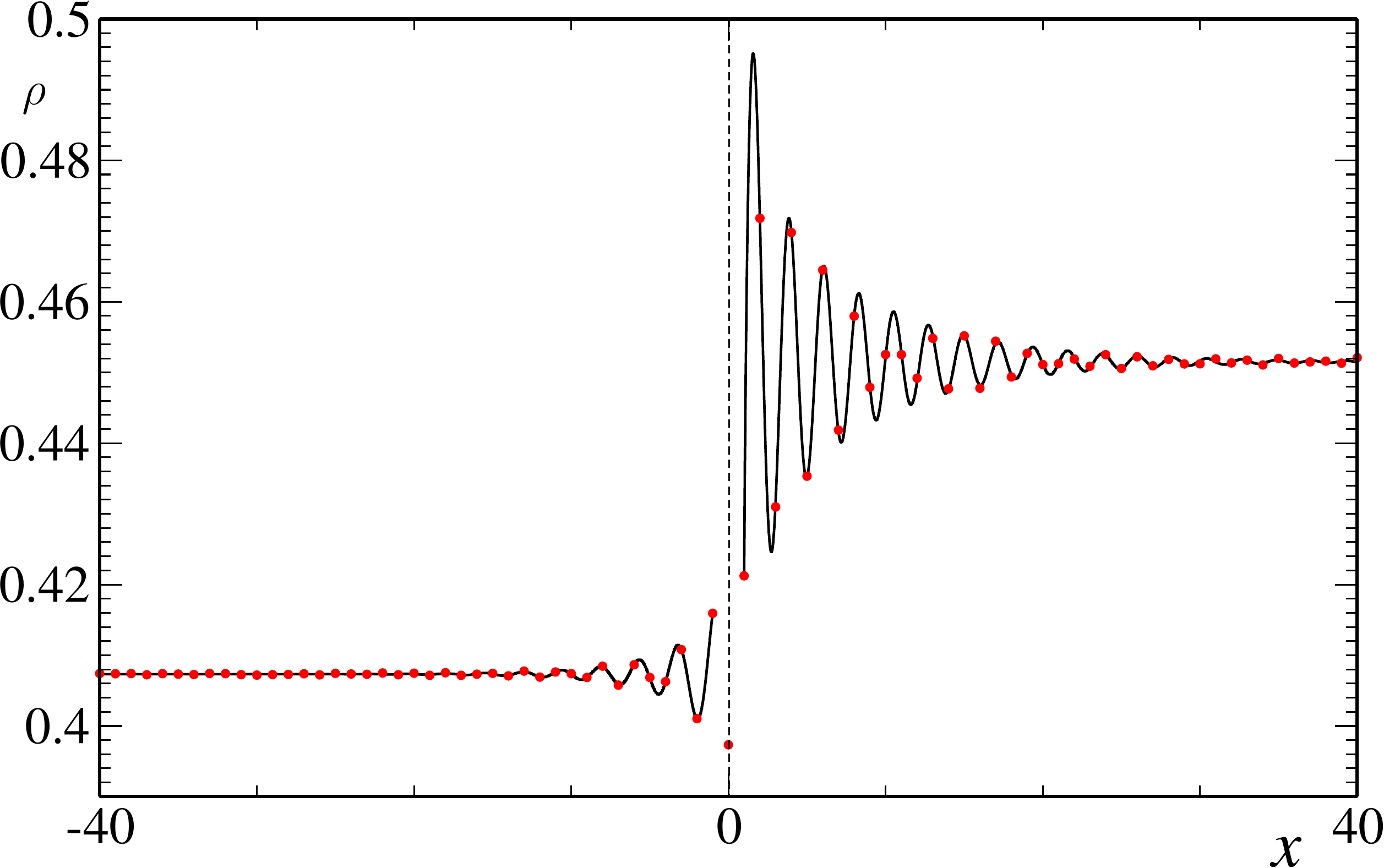}
\caption{(Color online) The full density including Friedel oscillations near the boundary, numerical results (filled circles) are fitted to the analytical result of Eq.~\eqref{dense} (lines) with  $t_l=1.31 t_r$, $U_\ell=0$, and  $U_r=1.8t_r$. The local potential energy is $V=0.75t_r$
and $t_r \beta=10$.}
\label{density_fig2}
\end{figure}

The local compressibility is defined as
\begin{equation}
\chi_x=-\frac{\partial\langle \hat n_x\rangle}{\partial \delta V}\bigg|_{\delta V=0}\,,
\end{equation}
analogous to the local susceptibility in a spin chain.\cite{Eggert1995} For the alternating contribution this yields
\begin{equation}
\chi_{\rm alt}\propto \lambda x\tau(x)\cos[2k_{F,x}^*x]\,.
\end{equation}
Unlike the Friedel oscillations in the density this observable remains non-zero even for half-filling and is therefore in that
particular case a more useful quantity to study.}

\subsection{Conducting fixed points}

In Sec.~\ref{sec_abrupt} we have predicted that for the abrupt
junction considered only one parameter needs to be tuned in order to
find a conducting fixed point.  The low-order expansion for $\lambda$
given by Eq.~\eqref{lambda} is not sufficient however to find the
location of the fixed points for the large interaction strengths we
want to consider in general.  Only in the limit $U_x\to 0$, where we
know the exact result, can we be confident of its predictions. An
exception is the half-filled case where we have previously
argued\cite{Sedlmayr2012a} that the scattering amplitude $\lambda$
vanishes for all interaction strengths if $u_\ell=u_r$, with the
velocities at half-filling known in closed form as a function of the
interaction strength from Bethe ansatz.\cite{tak99,HubbardBook}

\begin{figure}
\includegraphics*[width=0.99\columnwidth]{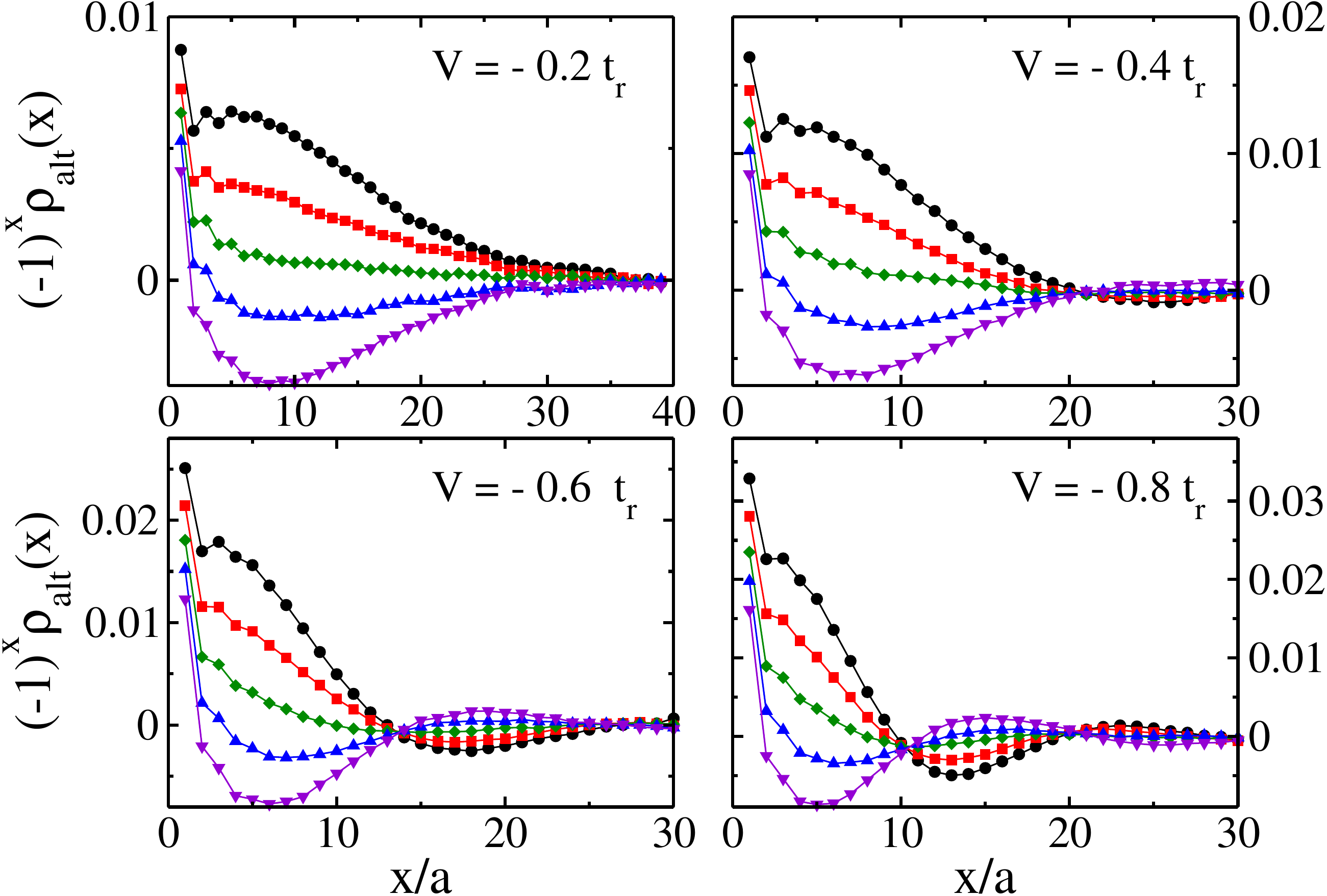}
\caption{(Color online) Plotted are the Friedel oscillations for different local potential energies $V$ calculated by QMC simulations, see main text for details, on the right hand side of the junction ($x>0$).  We only show the longer wavelength amplitude of the
rapid oscillations. In each panel from top to bottom: $t_\ell=1.3t_r$ for (black) circles, $t_\ell=1.4t_r$ for (red) squares, $t_\ell=1.5t_r$ for (green) diamonds, $t_\ell=1.6t_r$ for (blue) up-triangles, and $t_\ell=1.7t_r$ for (purple) down-triangles. We have used everywhere $U_\ell=0$, $U_r=1.8t_r$ and inverse temperature $t_r\beta=10$.}
\label{Fixed_Point}
\end{figure}

Instead, at generic fillings, we can find the locations of the solutions $t^*(V)$ which
solve $\lambda(t_\ell=t^*,V)=0$, keeping $U_x$ and $t_r$ fixed, by
analyzing the local density or compressibility of the system by QMC simulations
described in the preceding subsection.
We find that,
away from half-filling, these {\it do not} correspond to $u_\ell =u_r$.
For $\lambda=0$ the density is determined entirely by the Hamiltonian
Eq.~\eqref{spinless_ham}, plus irrelevant perturbations. For
$\lambda\neq0$, on the other hand, the relevant backscattering term
contributes. By plotting the density for different $t_\ell$ in
Fig.~\ref{Fixed_Point} we can find the places where the leading corrections
vanish and $\lambda$ changes sign,\cite{Sedlmayr2012a} which typically
can be observed in the range $5a \alt  x \alt 10a$.
Since we can always identify a value of hopping where the leading contribution vanishes,
there must be a line of conducting fixed points in parameter
space. In turn the existence of a full line of fixed points demonstrates that
there is only one condition for the conducting fixed point,
$\lambda=0$ with real $\lambda$. We want to stress though that even at such a point in
parameter space there are still irrelevant backscattering processes
present which only vanish in the zero temperature limit
$\beta\to\infty$.

\section{Conformally invariant boundary theory}\label{sec_cft}

In the preceding sections it has been demonstrated that it is possible to find an unstable conducting fixed point in two wires connected at a junction by appropriately tuning the bulk parameters of the wires.
The existence of this fixed point
 immediately invites the question of the nature of the effective low energy theory.
Obviously translational invariance is lost and it is also not possible
to use mirror charges as would be the case for an open boundary condition.
Therefore it is highly non-trivial to postulate a
description in terms of a conformally invariant
theory in this case.  Nonetheless, as we will show in this section it is possible to
characterize this fixed point in terms of mode expansions and {\it two}
effective boundary Luttinger liquid parameters.
Particular attention is paid to the case of half-filling where we can
pinpoint the fixed point precisely. This allows convenient numerical checks of the results.

\subsection{Mode expansion and finite size spectrum}

In the absence of backscattering at a junction we have the bosonic Hamiltonian \cite{Maslov1995,Furusaki1996,Sedlmayr2012a}
\begin{equation}\label{spinless_ham0}
\hh=\int\ud x \frac{1}{2}\left(\frac{1}{g_x}(\partial_x\phi)^2+g_x(\partial_x\tphi)^2\right)
\end{equation}
Compared to \eqref{spinless_ham} the position, $x$, was rescaled on the
two sides of the junction such that $u_\ell,u_r\to 1$. The fields obey
the canonical commutation relation:
$[\phi(x),\partial_y\tphi(y)]=\im\delta(x-y)$. Therefore we have the
relation
\begin{eqnarray}
\partial_t\phi(x)&=&\im[H,\phi(x)]\\\nonumber&=&
g_x\partial_x\tphi(x)\,.
\end{eqnarray}

The corresponding Green's function can be determined from Eq.~\eqref{spinless_ham0}, see Eq.~\eqref{green}.  Here we explore other properties of this boundary condition. We are interested in the solutions of the classical equation of motion,
\begin{equation}\label{classicalem}
\left[\partial_t^2-g_x\partial_x\left(\frac{1}{g_x}\partial_x\right)\right]\phi(x,t)=0
\end{equation}
on a ring with circumference $2L$ where
\begin{eqnarray}
g_x=\left\{\begin{array}{ll}
g_\ell& \textrm{ if } -L<x<0\\
g_r& \textrm{ if } 0<x<L\,.
\end{array}\right.
\end{eqnarray}
At the boundaries $\phi(x)$ and $\partial_x\phi(x)/g_x$ have to be
continuous leading to the boundary conditions
\begin{eqnarray}
\label{bc1}
\phi(0^-)=\phi(0^+),&\quad & \phi(-L)=\phi(L)\\*[0.1cm]
\frac{\partial_x\phi(0^-)}{g_\ell}=\frac{\partial_x\phi(0^+)}{g_r}, &\quad & \frac{\partial_x\phi(-L)}{g_\ell}=\frac{\partial_x\phi(L)}{g_r} \nonumber
\end{eqnarray}
The classical equation of motion \eqref{classicalem} has oscillatory
solutions as well as solutions linear in $x$, see Appendix
\ref{app_mode} for details. We may expand the field $\phi(x)$ in these
solutions, while respecting the canonical commutation relation
\begin{eqnarray}
[\phi(x),\partial_t\phi (y)]=ig_x\delta (x-y)\,.
\end{eqnarray}
This leads to
\begin{widetext}
\begin{eqnarray}\label{mode2}
\phi (x,t)&=&
\phi_0+\frac{\bg\Pi t}{2L}+\frac{Qxg_{x}}{2\tg L}+\sum_{l=1}^\infty\left[ \frac{\e^{-\im \pi lt/L}}{ \sqrt{2\pi l}}
\left[\sqrt{\bg} \cos (\pi lx/L )a_{e,l}+
\frac{g_x}{\sqrt{\tg}}\im\sin (\pi lx/L )a_{o,l}\right] +\textrm{H.c.}\right]\,,\\\nonumber
\tphi (x,t)&=&
\tphi_0+\frac{1}{\tg}\frac{Q t}{2L}+\frac{\bg \Pi x}{2g_{x}L}-\sum_{l=1}^\infty\left[ \frac{\e^{-\im \pi lt/L}}{ \sqrt{2\pi l}}
\left[\frac{\sqrt{\bg}}{g_x} \im\sin (\pi lx/L )a_{e,l}+
\frac{1}{\sqrt{\tg}}\cos (\pi lx/L )a_{o,l}\right] +\textrm{H.c.}\right]\,.
\end{eqnarray}
\end{widetext}
As before we have the boundary Luttinger parameter
\begin{eqnarray}
\frac{1}{\bg}=\frac{1}{2}\left[\frac{1}{g_\ell}+\frac{1}{g_r}\right]\,,
\end{eqnarray}
which describes the conductance.\cite{Safi1995,Furusaki1996,Sedlmayr2012a}
Interestingly, we find in addition a second boundary Luttinger
parameter
\begin{eqnarray}
\tg=\frac{1}{2}\left[g_\ell+g_r\right]\,,
\end{eqnarray}
which is important for other correlation functions as we will
see below.  $\Pi$ is the field conjugate to $\phi_0$ with
$[\phi_0,\Pi]=i$. As this field is periodic,
$\phi_0\to\phi_0+\sqrt{\pi}$, it is clear that the eigenvalues of the
conjugate field $\Pi$ must be $2\sqrt{\pi}m$, where $m$ is an integer.
$Q$ is the field conjugate to $\tphi_0$ and
$\tphi_0\to\tphi_0+\sqrt{4\pi}$ so that the eigenvalues of the
conjugate field $Q$ are $\sqrt{\pi}n$ for integer $n$.

The classical equation of motion \eqref{classicalem} has to follow
from a classical least action principle from which the classical
Hamiltonian
\begin{equation}
H=\int_{0}^{2L} \frac{\ud x}{2g_x}\left[\left(\partial_t\phi\right)^2+\left(\partial_x\phi\right)^2\right]\,.
\end{equation}
is determined. Substituting the mode expansion into the Hamiltonian,
we may read off the finite size spectrum
\begin{equation}\label{finitespec}
E=\frac{\pi}{ L}\left[-\frac{1}{ 12}+ \frac{n^2}{4\tg}+m^2\bg+\sum_{l=1}^\infty l(m_{e,l}+m_{o,l})\right]\,.
\end{equation}
Here $n$ and $m$ are arbitrary integers while $m_{e/o,l}$ are
non-negative integers corresponding to the eigenvalues of
$a^\dagger_{e/o,l}a_{e/o,l}$. We have included the universal term in
the ground state energy $-c\pi /(12L )$ with $c=1$ for a periodic
system of length $2L$.

\subsection{Scaling properties of the conducting fixed point}

As usual, since we have imposed the same boundary condition
at both ends, we may read off the scaling dimensions of all single-valued
boundary operators in the bosonized theory from
the finite size spectrum. The scaling dimensions are
\begin{eqnarray}
\zeta_{m,n} = \frac{n^2}{4\tg}+m^2\bg
+\sum_{l=1}^\infty l(m_{e,l}+m_{o,l})\,.\label{scdims}
\end{eqnarray}
Each dimension corresponds to a different boundary operator. $m^2\bg$
corresponds to $\exp [im\sqrt{4\pi }\phi (0)]$ with the $m=\pm 1$ operators being
the leading relevant operators at the unstable fixed point.
$\tg/4$ is the dimension of the operators $\exp [\pm i\sqrt{\pi }\tphi (0)]$, which
effectively correspond to spin operators $S^{\pm}(x=0)$, see below.

\begin{figure}
\includegraphics*[width=0.99\columnwidth]{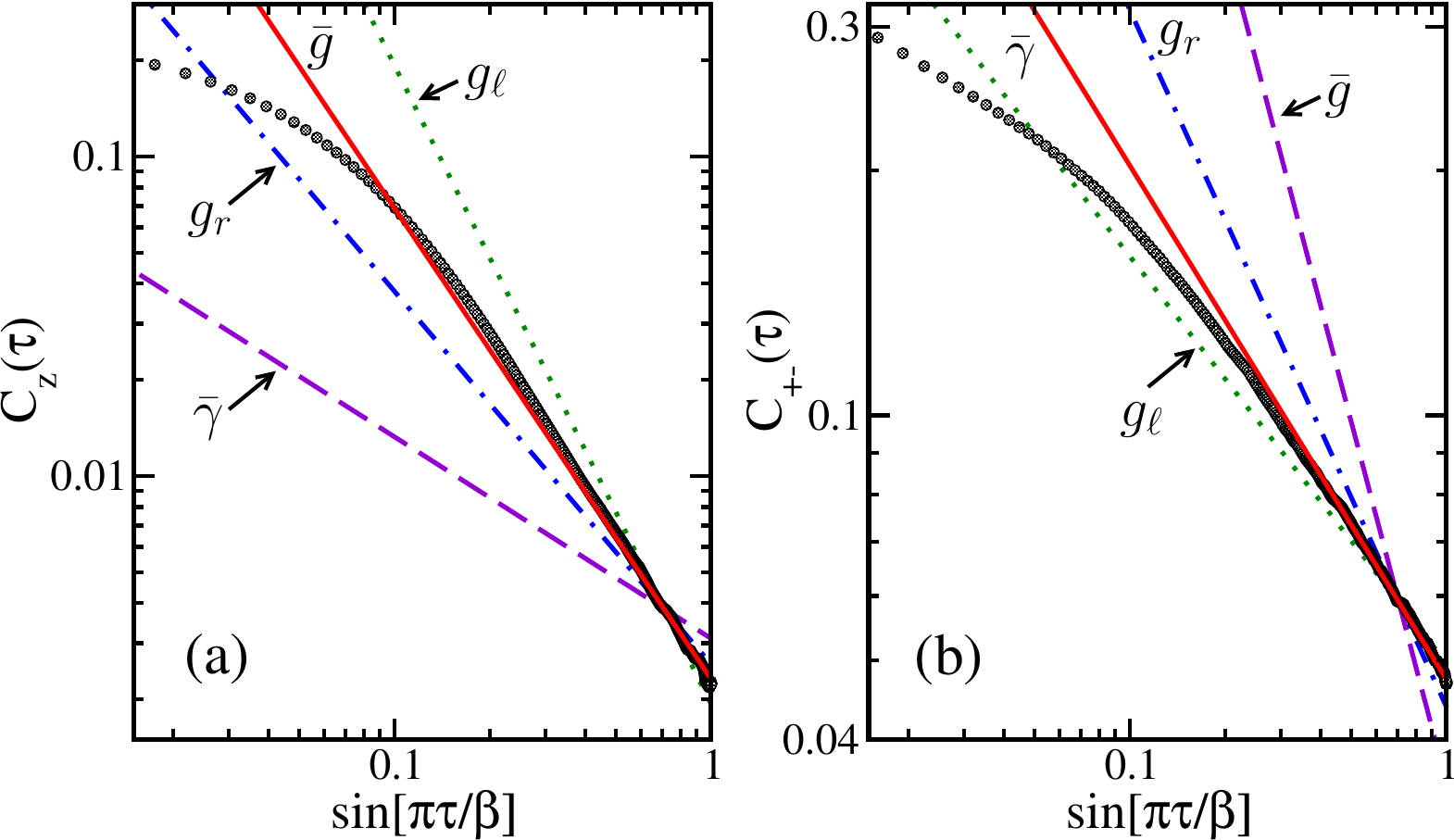}
\caption{(Color online) The scaling of the local spin correlation
  functions $C_z(\tau)$ and $C_{\pm}(\tau)$ at the fixed point:
  $t_\ell=1.518 t_r$, $U_\ell=0$, and $U_r=1.8t_r$. The magnetic field
  is zero (i.e.~$V=0$ for the corresponding fermion system) and the
  temperature is $t_r \beta=25$. (a) Numerical data, black circles,
  are compared to the predicted scaling $f(\tau )^\nu$ with
$f(\tau )\equiv |\sin (\pi \tau /\beta )|^{-2}$ and $\nu = \bar{g}$. (red curve).
   As a comparison we also
  plot, $f(\tau )^\nu$ with $\nu =g_\ell$, $g_r$, $\bar{\gamma}$, see Eq.~\eqref{cz}. (b)
  Numerical data, black circles, are compared to the predicted scaling
  $f(\tau )^{1/4\nu}$ with $\nu = \bar{\gamma}$ (red curve). As a
  comparison we also plot, $f(\tau )^{1/4\nu}$ with
$\nu =g_\ell$, $g_r$, $\bar{g}$, see
  Eq.~\eqref{cpm}.}
\label{Scaling_I}
\end{figure}
To analyze the scaling properties of the system, and compare the results with numerical calculations, it is convenient to introduce correlation functions for a spin system equivalent to our fermionic system.
The mapping between spin operators and fermionic operators is
given by the Jordan-Wigner transformation
\begin{equation}
S^+_j=\dpsi_j\e^{\im \pi\sum_{l<j}\dpsi_l\psi_l}\,.
\end{equation}
The leading $S^+S^-$ correlation function at the boundary $x=0$ is, in bosonized form,
\begin{equation}
\langle S^+(0,t)S^-(0,0)\rangle\sim
\left\langle \e^{-\im \sqrt{\pi}\tphi(0,t)}
\e^{-\im \sqrt{\pi}\tphi(0,0)}
\right\rangle\,.
\end{equation}
Using Eq.~\eqref{mode2} this results in
\begin{equation}
\langle S^+(0,t)S^-(0,0)\rangle\sim 
\left|\frac{\sin[\pi t/2L]}{\sin[\pi a/2L]}\right|^{-\frac{1}{2\tg}}\,,
\end{equation}
with the boundary exponent $\tg$.
For the $S^z$ operator we have after bosonization
\begin{equation}
S^z_j=\dpsi_j\psi_j-\frac{1}{2}= -\frac{a}{\sqrt{\pi}}\partial_x\phi +(-1)^j\mbox{const}\times\sin[\sqrt{4\pi}\phi]\,.
\end{equation}
The leading
$S^z$ spin density waves are described by the autocorrelation function at the boundary
\begin{equation}
\langle S^z(0,t)S^z(0,0)\rangle\sim 
\sum_{\alpha=\pm}\left\langle e^{\im\alpha\sqrt{4\pi}[\phi(0,t)-\phi(0,0)]} \right\rangle\,.
\end{equation}
From this one finds
\begin{equation}
\langle S^z(0,t)S^z(0,0)\rangle\sim
\left|\frac{\sin[\pi t/2L]}{\sin[\pi a/2L]}\right|^{-2\bg}\,,
\end{equation}
with the boundary exponent $\bg$. Thus the boundary theory is
described by {\it two different} boundary Luttinger parameters, $\bg$
and $\tg$.

In the QMC simulations we consider finite temperatures in the limit of
large system sizes $L \gg u\beta$ and calculate the imaginary time
correlation functions.
In this case the results are most easily accessible by considering the Green's function Eq.~\eqref{green}, and the equivalent correlation function for the adjoint field $\tphi(x,\tau)$. Then we find
\begin{equation}\label{cpm}
C_{\pm}(\tau)\equiv\langle S^+(0,\tau)S^-(0,0)\rangle 
\sim\left|\frac{\sin[\pi \tau/\beta]}{\pi a/\beta}\right|^{-\frac{1}{2\bar{\gamma}}}\,,
\end{equation}
and
\begin{equation}\label{cz}
C_z(\tau)\equiv\langle S^z(0,\tau)S^z(0,0)\rangle  
\sim\left|\frac{\sin[\pi \tau/\beta]}{\pi a/\beta}\right|^{-2\bar{g}}\,.
\end{equation}

We compare the predicted scaling of these correlation functions with the results of
QMC simulations. The predicted exponents are well verified, see Fig.~\ref{Scaling_I}. Not only can one clearly distinguish the two boundary exponents, but we have also checked that the bulk exponents do not fit the scaling. Note that the analytical formula are only valid in the asymptotic limit $\tau \gg \beta$. The values of $g_{\ell,r}$, and hence of $\tg$ and $\bg$, can be found exactly from the results of Bethe ansatz.\cite{tak99,HubbardBook,PereiraSirkerJSTAT,SirkerLL}

\begin{figure}
\includegraphics*[angle=270,width=0.99\columnwidth]{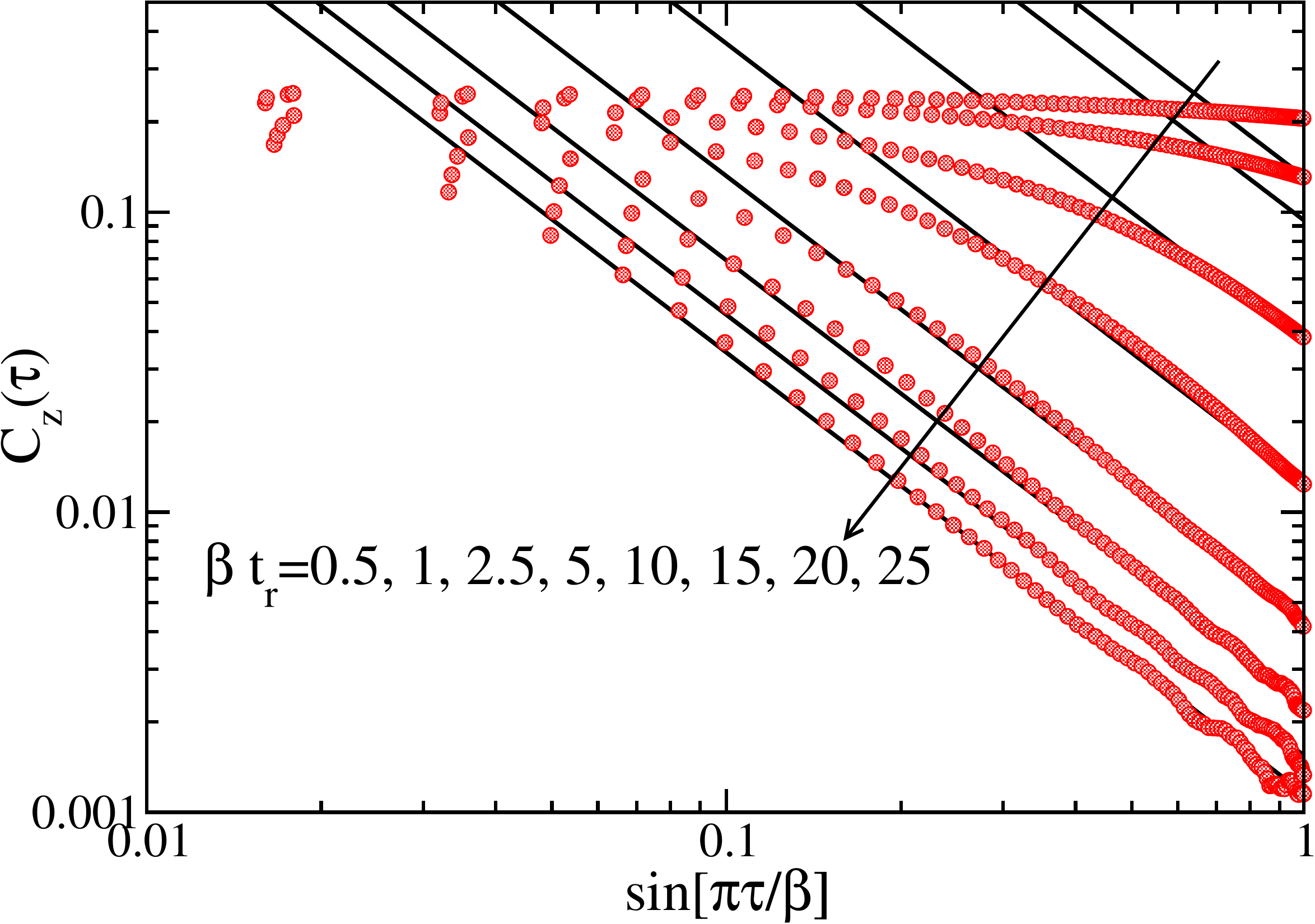}
\caption{(Color online) The scaling of the local spin correlation
  function $C_z(\tau)$ at the conducting fixed point with
  $t_\ell=1.518t_r$, $U_\ell=0$, and $U_r=1.8t_r$. The magnetic field
  is zero (i.e.~$V=0$ for the corresponding fermion system). Numerical
  data for different inverse temperatures (as indicated on the plot)
  are compared to the predicted scaling (lines). As temperature is
  lowered the field theory becomes more accurate.}
\label{Fixed_Point_SZSZ_Scaling}
\end{figure}
\begin{figure}
\includegraphics*[angle=270,width=0.99\columnwidth]{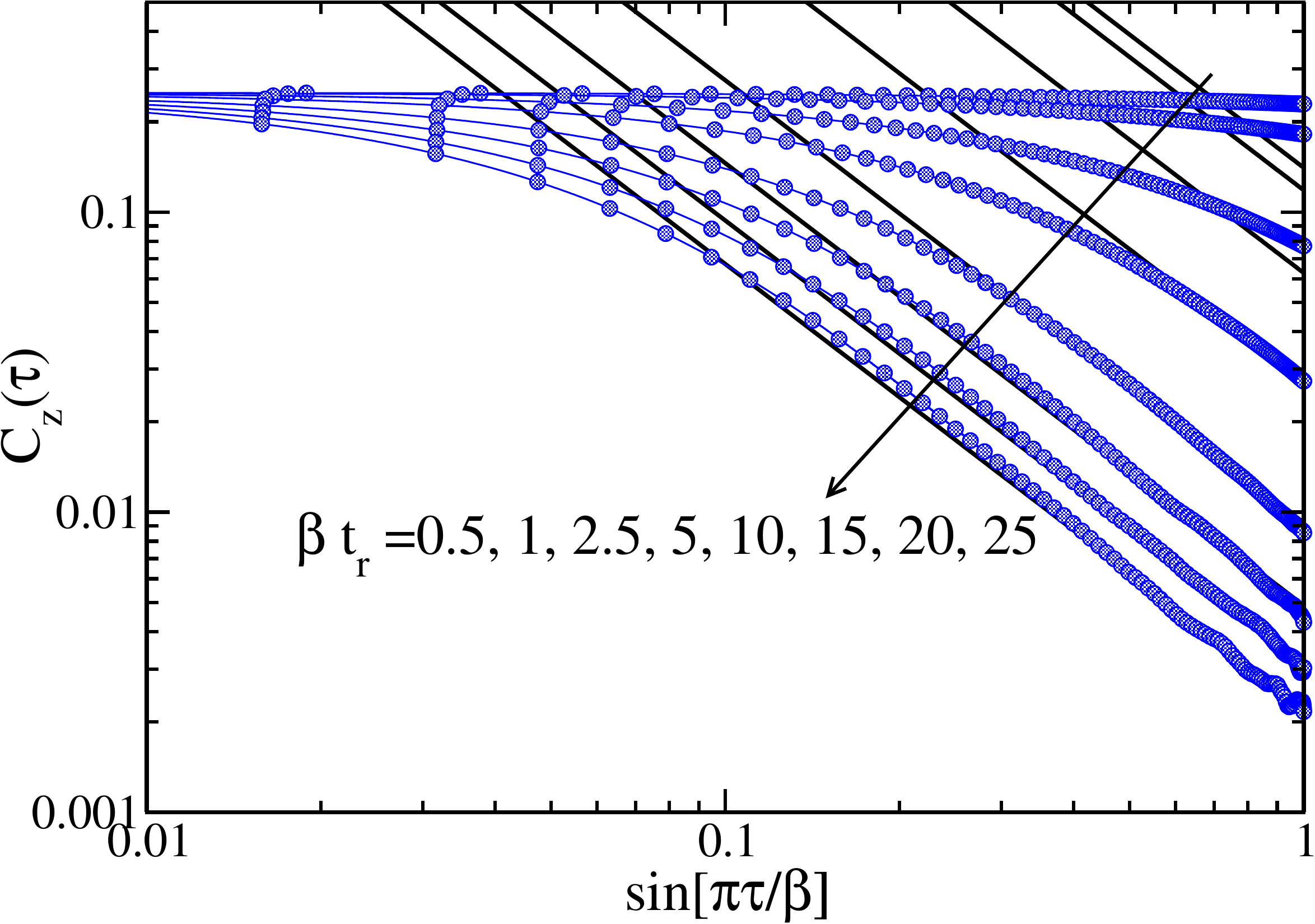}
\caption{(Color online) The scaling of the local spin correlation
  function $C_z(\tau)$ away from the conducting fixed point with
  $t_\ell=t_r$, $U_\ell=0$, and $U_r=1.8t_r$. The magnetic field is
  zero (i.e.~$V=0$ for the corresponding fermion system). From top to
  bottom we plot different values of the inverse temperature $t_r
  \beta= \{0.5, 2.5, 5, 10, 15, 20, 25\}$.}
\label{SZSZ_Scaling}
\end{figure}

Fig.~\ref{Fixed_Point_SZSZ_Scaling} shows the temperature scaling of
$C_z(\tau)$ at the conducting fixed point $t_\ell=1.518t_r$, and
Fig.~\ref{SZSZ_Scaling} the scaling away from it.
At the conducting fixed point the field theory, as expected, does not
describe the data at high temperatures, such as $t_r\beta=0.5$ or
$t_r\beta=2.5$. As temperature is lowered the field theory becomes a better
and better fit, showing good scaling already by $t_r\beta=5$.  As we
move away from the conducting fixed point the corrections to scaling
are expected to grow while lowering the temperature but are only
$\Order(\lambda^2)$. This makes it impossible to see the approach to
the insulating fixed point in $C_z(\tau)$. The Friedel oscillations of
the density and compressibility considered in Sec.~\ref{sec_density}
are, in principle, better to see the crossover to the insulating fixed
point. However, the expected cross-over temperature is of order
$T\approx10^{-4}t_r$,\cite{Sedlmayr2012a} which is unfortunately well
beyond the reach of the numerical QMC simulations.

\subsection{Local density of states}

One possible experimental test on boundary exponents is the measurement of the local
density of states with local spectroscopic tools, such as scanning tunneling
spectroscopy.\cite{Blumenstein11}  Theoretically a characteristic depletion
with the boundary exponent has been predicted,\cite{Kane1992,Eggert1996,Eggert00,Kakashvili06,Anfuso03,Schneider10} which may be corrected by
irrelevant operators.\cite{Soeffing2013}  It is therefore interesting to calculate the
characteristic signatures of the local density of states for this unusual
fixed point.

The local density of states is defined as
\begin{eqnarray}
\nu(x,\omega)&=&\frac{1}{\pi}\int\ud te^{\im \omega t}{\rm Re}\langle\psi(x,t)\dpsi(x,0)\rangle\\\nonumber&=&
\frac{1}{2\pi^2a}\int\ud t\e^{\im \omega t}\sum_{\alpha=\pm} e^{-2\pi\langle(\phi_\alpha(x,t)-\phi_\alpha(x,0))^2\rangle}
\end{eqnarray}
Using the correlation functions calculated from the mode expansion, see Appendix \ref{app_mode}, and, neglecting the cut-off for the moment, this results in
\begin{eqnarray}\label{dos}
\nu(x,\omega)&\sim&\frac{1}{2\pi^2a}\int\ud t\e^{\im \omega t}\left|4\sin^2[\pi t/2L]\right|^{-\delta_x} \\\nonumber&&\times
\left|4\sin[\pi (t-2x)/2L]\sin[\pi (t+2x)/2L]\right|^{-\kappa_x}\\\nonumber&&\times
\sum_{\alpha=\pm}\left|\frac{\sin[\pi (t-2x)/2L]}{\sin[\pi (t+2x)/2L]}\right|^{\alpha/2}\,.
\end{eqnarray}
The exponents are given by
\begin{eqnarray}
\kappa_x&=&\frac{1}{8}\left[\bg+\frac{1}{\tg}-\frac{\bg}{g_x^2}-\frac{g_x^2}{\tg}\right]=\frac{1}{4}\frac{g_x-g_{- x}}{g_\ell+g_r}\left(\frac{1}{g_x}-g_x\right)\,,\nonumber\\
\delta_x&=&\frac{1}{8}\left[\bg+\frac{1}{\tg}+\frac{\bg}{g_x^2}+\frac{g_x^2}{\tg}\right]=\frac{1}{4}\left(\frac{1}{g_x}+g_x\right)\,.
\end{eqnarray}
In the bulk regions near $x\approx L/2$ we recover $\nu(0,\omega)\sim|\omega|^{\xi_x}$ with the usual exponents $\xi_x=2\delta_x-1$.

The local density of states at the boundary $x=0$ therefore becomes, reinstating a cut-off of the order of the lattice spacing $a$,
\begin{equation}
\nu(0,\omega)\sim\frac{1}{\pi^2a}\int\ud t\e^{\im \omega t}
\left|\frac{\sin[\pi t/2L]}{\pi a/2L}\right|^{-2\zeta}
\end{equation}
giving $\nu(0,\omega)\sim|\omega|^{2\zeta -1}$ with scaling dimension
\begin{equation}
\zeta \equiv\frac{1}{4}\left[\bg+\frac{1}{\tg}\right]=\frac{1+g_\ell g_r}{2(g_\ell+g_r)}\,.
\end{equation}
Note that this is {\it not} one of the dimensions of single-valued operators, $\exp [i\sqrt{\pi}(n\tilde \phi +2m\phi )]$ for
integer $n$, $m$ listed in Eq. (\ref{scdims}).  Rather $\psi_\pm \propto \exp [i\sqrt{\pi}(\pm \tilde \phi +\phi )]$, corresponding
to $n=\pm 1$, $m=1/2$.  The non single-valued nature of these operators is a result of their being fermonic.
$\zeta$ is the same as the bulk scaling dimension in a homogeneous spinless Luttinger liquid with $g\to\bg$ and $g^{-1}\to\tg^{-1}$. Surprisingly, the density of states at the junction scales as in the free fermion case if either side of the junction is non-interacting: $g_\ell=1$ or $g_r=1$. This can be understood from the density of states, Eq.~\eqref{dos}, if we have $g_\ell=1$ then the exponent $\kappa_{x<0}=0$. Both in the vicinity of the boundary and in the bulk the last line in Eq.~\eqref{dos} does not affect the scaling properties of the density of states. Hence the scaling of $\nu(x<0,\omega)$ on the non-interacting side is no longer position dependent and shows the bulk scaling right up to the boundary itself. This is not true on the interacting side ($x>0$) where the scaling modulates from the non-interacting result at the boundary $x=0$, to the bulk interacting value far inside the wire. In contrast to the density or compressibility of Sec.~\ref{sec_density} there is no proximity effect near the boundary in the non-interacting wire.

\subsection{Fixed points and the g-theorem}

From the finite size spectrum, Eq.~\eqref{finitespec}, we may also read off the partition function in the scaling limit:
\begin{equation}
Z(\beta/L)=\eta^{-2}\left( \e^{-\pi \beta /L }\right)
\theta_3\left(\e^{-\pi \beta /[2\tg L]}\right)
\theta_3\left(\e^{-\pi \beta 2\bg/L}\right)\,.
\end{equation}
Here we have introduced the Dedekind eta and Jacobi theta functions,
\begin{eqnarray}
\eta (q)&\equiv& q^{1/24}\prod_{n=1}^\infty (1-q^n)\textrm{ and}\nonumber \\
\theta_3(q)&\equiv&\sum_{n=-\infty}^\infty q^{n^2/2}\,.
\end{eqnarray}
In the thermodynamic limit, $\beta /L \to 0$, this becomes
\begin{equation}
Z\to \sqrt{\tg/\bg}\e^{\pi L /(3\beta )}\,.
\end{equation}
Apart from the usual bulk free energy, $F=-\pi L /(3\beta^2)$, there is
also a ``ground state degeneracy'', $g_d$, associated with the two interfaces in the system.
The factor for each interface is
\begin{equation}
g_d^c=\left(\frac{\tg}{\bg}\right)^{\frac{1}{4}}=\frac{(g_\ell+g_r)^{1/2}}{(4g_\ell g_r)^{1/4}}\,.
\end{equation}
This may be compared to the ground state degeneracy for the insulating fixed point
where the junction consists of the perfectly reflecting ends of two quantum wires
with Luttinger parameters $g_\ell$ and $g_r$. This fixed point has \cite{Eggert1992}
\begin{equation}
g_d^i=(g_\ell g_r)^{1/4}\,,
\end{equation}
According to the ``g-theorem'',
boundary RG flows between fixed points can only
occur when $g_d$ is reduced during the flow.\cite{Affleck1991}  Therefore, it is interesting to consider the ratio
\begin{equation}
g_d^c/g_d^i=\frac{1}{\sqrt{\bg}}\,.
\end{equation}
The g-theorem states that flow from the conducting to insulating fixed point is only possible when $\bg<1$. This is
consistent with the analysis here since $\bg$ is the dimension of the operator
$\cos[\sqrt{4\pi}\phi(0)]$ which drives the flow. The flow only takes place when the operator
is relevant, corresponding to $\bg<1$.  For sufficiently large $g^i_d$ the renormalization flow can
occur from insulating to conducting fixed points. As shown in Ref.~\onlinecite{Eggert1992},
the fermion operator, or equivalently spin raising operator, at the end of the open chain,
has scaling dimension $1/(2g_\ell)$ or $1/(2g_r)$ as appropriate. We might expect the flow from insulating to conducting when
the tunneling between the two open chains is relevant which occurs when
\begin{equation}
\frac{1}{2g_\ell}+\frac{1}{2g_r}=\frac{1}{\bg}<1
\end{equation}
and hence $\bg>1$.
In this case $g^c_d<g^i_d$ so this flow is also consistent with the g-theorem.

It is also interesting to consider the flow starting from the
insulating fixed point, but with a weakly connected resonant site in between the two wires: the resonant fixed point. Then for a range
of Luttinger parameters an RG flow from the resonant to the conducting fixed point is expected. A necessary condition for the flow from
resonant to conducting fixed points is that the tunneling operators from each chain to the
resonant site are relevant, $g_{\ell,r}>1/2$.
The ground state degeneracy of the resonant fixed point is bigger by a factor of 2 than that of
the insulating fixed point due to the 2-fold degeneracy of the resonant site and
\begin{equation}
g_d^r=2(g_\ell g_r)^{1/4}\,.
\end{equation}
Thus the ratio of ground state degeneracies of the resonant to conducting fixed points is
\begin{equation}
g_d^r/g_d^c=2\sqrt{\bg}\,.
\end{equation}
We can see that
$g_d^r/g_d^c\geq \sqrt{2}$ whenever $g_\ell,g_r>1/2$ so the g-theorem is also obeyed by this RG flow.
Even when $\lambda$ is tuned to zero, corresponding to resonance, the next most relevant operators, $\exp [\pm 2i\sqrt{4\pi}\phi (0)]$
will still be present.  This can drive the flow from the conducting to the resonant fixed points when it becomes relevant, \emph{i.e.}~for $4\bg<1$.
Since $g_d^c/g_d^r=1/(2\sqrt{\bg})$ we see that this flow is consistent with the g-theorem as it
only occurs when $\bg <1/4$. Therefore all expected RG flows are consistent with the g-theorem.

As first observed by Kane and Fisher\cite{Kane1992} in the case $g_\ell=g_r$, there is a range of Luttinger parameters where
both conducting and resonant fixed points are stable. In this case they are separated by an intermediate
unstable fixed point.

\section{Conclusions}\label{conclusions}
In conclusion, we have described a novel conducting fixed point in
inhomogeneous quantum wires. This fixed point is reached by tuning to
zero the amplitude of the leading backscattering operator at the
junction between two homogeneous parts of the wire. We have, in
particular, studied a lattice model of spinless fermions with nearest
neighbor hopping and interaction in the
critical regime. For the case of an abrupt junction we have derived
the backscattering amplitude for all fillings in lowest order in the
interaction. For the half-filled case it is even possible to give a
condition for the vanishing of the backscattering amplitude valid for
all interaction strengths. The prediction of a conducting fixed point
were numerically confirmed by numerical QMC calculations of the
Friedel oscillations in the local density and compressibility close to
the boundary which vanish in leading order at the fixed point.

One of our main results is the derivation of the boundary conformal field for
this novel unstable conducting fixed point. The conformally invariant
theory for this case is highly unusual because the two parts of the
wire are governed by different bulk Luttinger parameters $g_\ell$ and
$g_r$. As a consequence, we find that the scaling dimensions of
boundary operators are also governed by two different Luttinger
parameters given by $\tg=(g_\ell+g_r)/2$ and $\bg=2g_\ell
g_r/(g_\ell+g_r)$. We showed, both analytically and numerically, that
$\tg$ is controlling the transverse spin autocorrelation function
while $\bg$ controls the longitudinal one in the corresponding spin
model. Experimentally, a test of the boundary exponents could possibly
be obtained by scanning tunneling microscopy which would allow one to
measure the local density of states which shows energy scaling with an
exponent being determined by $\tg$ and $\bg$.

\acknowledgments J.S., S.E., and N.S. acknowledge support by the
Collaborative Research Center SFB/TR49 and the graduate
school of excellence MAINZ.
The research of I.A. was supported by NSERC and CIfAR.
We are grateful for computation time at AHRP.

\appendix

\section{Landauer formula for transmission}\label{App_Landauer}

The significance of the measure of transmission $R=0$ can be verified by considering the Landauer formula.\cite{Landauer1957}
Thus we imagine attaching the wire to reservoirs on the left and right side with different chemical
potentials $\mu_L$ and $\mu_R$.  We consider particles emitted from the left reservoir with a thermal distribution with
chemical potential $\mu_L=-eV_L$ and from the right reservoir with a thermal distribution and chemical potential $\mu_R=-eV_R$.
At zero temperature the Fermi wave-vectors on left and right sides, $k_{F\ell}$ and $k_{Fr}$, are given by
\begin{equation}
-2t\cos k_{F\ell,r}-V_{\ell,r} =\mu_{L,R}\,.\label{res}
\end{equation}
 Suppose that the bottom of the band on the left has higher energy than the bottom of the band on the right.
Then the total current, at zero temperature, is
\begin{eqnarray}
I&=&-e\int_0^{k_{F\ell}}\frac{\textrm{d}k}{2\pi}u_\ell (k)[1-|R(k)|^2]\\\nonumber&&+e\int_{-k_{Fr}}^{-k_{2max}}\frac{\textrm{d}k}{2\pi}u_\ell (k)|T(k)|^2\,.
\end{eqnarray}
The first term is the current emitted by the left reservoir and partially reflected at the interface. The second term
is the current emitted by the right reservoir and partially transmitted. The maximum wave-vector for the
second integral, $k_{2max}>0$,  is given by $\epsilon_2(-k_{2max})=\epsilon_1(0)$ since lower energy incoming
particles from the right have zero transmission probability.

It is convenient to change integration variables to $\epsilon_1$ in the first integral and $\epsilon_2$ in the
second, giving:
\begin{eqnarray}
I&=&-e\int_{\epsilon_1(0)}^{\mu_L}\frac{\textrm{d}\epsilon_1}{2\pi}[1-|R(\epsilon_1)|^2]\\\nonumber&&
+e\int_{\epsilon_1(0)}^{\mu_R}\frac{\textrm{d}\epsilon_2}{2\pi}\frac{u_\ell (\epsilon_2)}{u_r(\epsilon_2)}|T(\epsilon_2)|^2
\end{eqnarray}
Since $|T|^2u_\ell /u_r=1-|R|^2$ this can be written as
\begin{equation}
I=-e\int_{\mu_R}^{\mu_L}\frac{\textrm{d}\epsilon}{2\pi}[1-|R(\epsilon )|^2]\,.\label {L1}
\end{equation}
Now taking the limit $\mu_L\to \mu_R\equiv \epsilon_F$, we find:
\begin{equation}
I\to \frac{e^2}{2\pi}[1-|R(\epsilon_F)|^2](V_L-V_R)\,.\label{L2}
\end{equation}
Hence the linear conductance is
\begin{equation}
G=\frac{\ud I}{\ud V}=\frac{e^2}{2\pi}[1-|R(\epsilon_F)|^2]\,.\label{L3}
\end{equation}
This is another way of seeing that $[1-|R|^2]$ is the suitable measure of the transmission
of the interface.

\section{Non-interacting calculations}
\label{App_NonInt}
In the main text, Sec.~\ref{sec_nonint} we have considered the
simplest possible junction, a jump between two homogeneous regions, in
the non-interacting case. Here we want to present calculations for more
general junctions to study the influence on the backscattering term.
\subsection{Abrupt junction with additional local variation}
The calculation of Sec.~\ref{sec_nonint} can be extended
straightforwardly to a more general model where the hopping amplitude
varies near the origin.  Suppose, for example, that the hopping
amplitude from site -1 to 0 is $t_{-1}=t_\ell'$ and from 0 to 1 is
$t_0=t_r'$ with the rest as given by Eq.~\eqref{abrupt}. For
simplicity we concentrate again on half-filling, $V_j=0$. Then we
may write the wave-function for an incoming wave from the left as in
Eq.~\eqref{scatter} with $j_\ell=-1$ and $j_r=1$.  $\psi_0$ is now a
free parameter.  Solving for the reflection amplitude as previously
gives
\begin{equation}\label{RT}
|R|^2=\frac{\left[\frac{t_\ell'^2}{t_\ell^2}u_\ell -\frac{t_r'^2}{t_r^2}u_r\right]^2+a^2\epsilon^2\left[2-\frac{t_\ell'^2}{t_\ell^2}-\frac{t_r'^2}{t_r^2}\right]^2}{ \left[\frac{t_\ell'^2}{t_\ell^2}u_\ell +\frac{t_r'^2}{t_r^2}u_r\right]^2+a^2\epsilon^2\left[2-\frac{t_\ell'^2}{t_\ell^2}-\frac{t_r'^2}{t_r^2}\right]^2}\,.
\end{equation}
Solving for $R=0$ one finds
\begin{eqnarray}
(t_\ell'/t_\ell,t_r'/t_r)=\sqrt{2}(\cos \theta ,\sin \theta )
\end{eqnarray}
with $\tan^2\theta = u_\ell /u_r$.  Thus maximal conductance can be
achieved for any choice of energy $\epsilon$, and thus any value of $u_\ell
/u_r$ that can occur as $\epsilon$ is varied.

We see that
the simple condition $u_\ell =u_r$ for perfect conductance is a
special result, which only holds for the ``abrupt junction''
considered in the main text.  In general, the condition $u_\ell =u_r$
can be regarded as removing the intrinsic scattering from a sharp jump
between bulk values of the hopping. Additional variation on top of
this will naturally result in scattering and an additional finetuning
is required to reach the conducting fixed point.

Note that the fact that two parameters, $t_\ell'$ and $t_r'$, need to
be adjusted to achieve perfect conductance, in general is in accord
with the renormalization group (RG) viewpoint.  For non-zero
energy $\epsilon$, particle-hole symmetry is broken so the scattering
amplitude $\lambda$ can be complex. In the special case $\epsilon =0$
where particle-hole symmetry holds there is only one condition for
perfect conductance $t_\ell'^2/t_\ell^2=t_r'^2/t_r^2$ and only one
parameter needs to be adjusted.

Now let us consider the case with non-trivial $t_i'$ within the narrow band approximation.
We use the parametrization
\begin{eqnarray}
t_\ell'&=&t_\ell+\delta t_\ell'=t-\delta t + \delta t_\ell'\nonumber \\
t_r'&=&t_r+\delta t_r'=t+\delta t+\delta t_r'\,.\label{t12'}
\end{eqnarray}
In this case, there is another backscattering perturbation term
\begin{eqnarray}
\delta \hh'
&=&-2\delta t_\ell'\psi^\dagger_-\psi_+\e^{-\im k_Fa}-2\delta t_r'\psi^\dagger_-\psi_+\e^{\im k_Fa}+\textrm{H.c.}\nonumber \\&&
-(\delta t_\ell'+\delta t_r')2\cos[k_Fa](\psi^\dagger_-\psi_-+\psi^\dagger_+\psi_+)
\end{eqnarray}
where $\psi_-$ and $\psi_+$ are evaluated at $x=0$ in all terms. Focusing on the backscattering term the perturbation becomes
\begin{equation}
\delta \hh+\delta \hh'=2\pi\im\lambda \psi^\dagger_-\psi_+(x=0)+\textrm{H.c.}
\end{equation}
with
\begin{equation}
\lambda=\frac{u_\ell-u_r}{4 \pi a}+\frac{\im\delta t_\ell'\e^{-\im k_Fa}+\im\delta t_r'\e^{\im k_Fa}}{\pi}\,.
\end{equation}
Although the variations in $t_i$ are small,
they occur over only three sites, so this is not an adiabatic change.
Note that while we were able to
determine $\lambda$ explicitly in this model, with all $t_i$ nearly equal, it
may not be feasible to do so in all cases. In fact, a reduction to a narrow
band model is not accurate in general, as discussed in the main text.

\subsection{Next-nearest neighbor hopping}
Next-nearest neighbor hopping can also be added to the Hamiltonian,
explicitly breaking particle-hole symmetry. We consider the
Hamiltonian
\begin{equation}
\hh_0'=\sum_i[-t_{1,j}\psi^\dagger_i\psi_{i+1}-t_{2,i}\psi^\dagger_i\psi_{i+2}+\textrm{H.c.}]\,.
\end{equation}
To keep things as simple as possible we choose
\begin{equation}
t_{1,i}=\left\{\begin{array}{ll} t_{1L},&(i\leq -1)\\ t_{1R},&(i\geq 0) \end{array}\right.\,,
\quad t_{2,i}=\left\{\begin{array}{ll} t_{2L},&(i\leq -2)\\ t_{2R},&(i\geq 0)\end{array} \right.\,.
\end{equation}
There is no particularly simple or natural choice for $t_{2,-1}$ so it is kept as a free parameter.
Let us assume that all the $t_{1,i}$ are close together and all the $t_{2,i}$ are close
together so that the narrow band approach is applicable. Thus we write
\begin{eqnarray}
t_{1L}&=&t_1-\delta t_1\,,\quad t_{2L}=t_2-\delta t_2\,,\nonumber \\
t_{1R}&=&t_1+\delta t_1\,,\quad t_{2R}=t_2+\delta t_2\,,\\
t_{2,-1}&=&t_2+\delta t\,.\nonumber
\end{eqnarray}
A simple extension of the previous calculation gives
\begin{equation}
\pi\lambda =-\delta t_1\csc[k_Fa]-\delta t_2\cot[k_Fa]+\im\delta t\e^{2\im k_Fa}
\end{equation}
for the back-scattering coupling constant.

As a simpler special case, consider $k_{F}=\pi /2$. Now
\begin{equation}
\pi\lambda = -\delta t_1-\delta t_2-\im\delta t\,,
\end{equation}
and $\lambda$ is complex in this case despite being at half-filling;
this is natural since $t_2$ breaks particle-hole symmetry at all
fillings.

\section{Bosonization details}\label{H_Appendix}

First we note the following useful relations:
\begin{eqnarray}
\dpsi_\alpha(x)\psi_{\alpha}(x)&=&\rho_\alpha(x)\equiv-\frac{1}{\sqrt{\pi}}\partial_x\phi_\alpha(x),\nonumber\\
\dpsi_\alpha(x)\partial_x\psi_\alpha(x)&=&\alpha i\pi\rho_\alpha^2(x),\textrm{ and}\\\nonumber
\dpsi_\alpha(x)\psi_{-\alpha}(x)&=&\frac{i\alpha}{2\pi a}e^{-i\alpha\sqrt{4\pi}\phi(x)}.
\end{eqnarray}

Due to the inhomogeneous nature of both the Fermi momentum and the
interactions, the $2k_{F,x}$ oscillating terms in the interaction can
no longer be neglected, see Eq.~\eqref{linear_int_ham}.
One finds a
nonzero contribution to the backscattering around any region of
inhomogeneity. These terms must be treated carefully, as an example we
can take
$:\dpsi_\alpha\psi_\alpha(x)::\dpsi_\alpha\psi_{-\alpha}(x+a):$.
Direct rearrangement gives for
$\dpsi_\alpha\psi_\alpha(x)\dpsi_\alpha\psi_{-\alpha}(x+a)$ either
\begin{eqnarray}
:\dpsi_\alpha\psi_\alpha(x)::\dpsi_\alpha\psi_{-\alpha}(x+a):\qquad\qquad\qquad\\\nonumber
+\langle 0|\dpsi_\alpha\psi_\alpha(x)|0\rangle:\dpsi_\alpha\psi_{-\alpha}(x+a):
\end{eqnarray}
or
\begin{eqnarray}
-:\dpsi_\alpha(x)\psi_{-\alpha}(x+a)::\dpsi_\alpha(x+a)\psi_\alpha(x):\qquad\qquad\quad\\\nonumber
\qquad\qquad-:\dpsi_\alpha(x)\psi_{-\alpha}(x+a):\langle 0|\dpsi_\alpha(x+a)\psi_\alpha(x)|0\rangle\,,
\end{eqnarray}
which are therefore equal.
Then, using
\begin{equation}
\langle 0|\dpsi_\alpha(x+a)\psi_\alpha(x)|0\rangle\approx \im\alpha/2\pi a\,,
\end{equation}
and expanding in the cut-off $a$ this allows us to write
\begin{eqnarray}
:\dpsi_\alpha\psi_\alpha(x)::\dpsi_\alpha\psi_{-\alpha}(x+a):\approx\hspace{3.2cm}\\\nonumber
-\langle 0|\dpsi_\alpha\psi_\alpha(x)|0\rangle:\dpsi_\alpha\psi_{-\alpha}(x+a):\hspace{2cm}\\\nonumber
-:\dpsi_\alpha(x)\psi_{-\alpha}(x+a):\hspace{2cm}\\\nonumber\times\left(\frac{\im\alpha}{2\pi a}+\rho_\alpha(x)+a\psi_\alpha^\dagger\partial_x\psi_\alpha(x)\right)\,.
\end{eqnarray}
Now keeping only the leading order terms we find
\begin{equation}
:\dpsi_\alpha\psi_\alpha(x)::\dpsi_\alpha\psi_{-\alpha}(x+a):\approx \frac{e^{-\im \alpha\sqrt{4\pi}\phi(x)}}{4\pi^2 a^2}\,.
\end{equation}
Similar expressions hold for the other terms.

The bosonized free and interacting Hamiltonians become
\begin{eqnarray}
\hh_0&=&-\sum_{x \alpha}\alpha it_xa^2e^{\im \alpha\kappa^-_x}(\partial_x\phi_\alpha)^2+\textrm{h.c.}\\&&
-\sum_{x \alpha}\frac{\im \alpha t_x}{\pi}e^{-2i\alpha k_{F,x}x-i\alpha\sqrt{4\pi}\phi(x)}\left[e^{-\im \alpha \kappa^-_x}-\frac{V}{2t_x}\right]\,,\nonumber
\end{eqnarray}
and
\begin{eqnarray}
\hh_I&=&\sum_{x \alpha}a^2U_x\bigg[\left(1-e^{-2i\alpha\kappa^-_x}\right)\frac{(\partial_x\phi)^2}{2\pi}\nonumber
\\&&\nonumber +e^{2i\alpha k_{F,x}x}\left(e^{2i\alpha \kappa^-_x}-1\right)\frac{2e^{\im \alpha\sqrt{4\pi}\phi(x)}}{(2\pi a)^2}\\&&
-e^{4i\alpha\kappa^+_x}\frac{e^{\im \alpha 2\sqrt{4\pi}\phi(x)}}{(2\pi a)^2}
\bigg]\,,
\end{eqnarray}
respectively. Included in this is the irrelevant umklapp scattering
\begin{equation}
\hh_U=-\sum_{x}a \frac{U_x}{\pi^2a}\cos\left[4\kappa^+_x+2\sqrt{4\pi}\phi(x)\right]\,.
\end{equation}
Away from half-filling this is only a boundary contribution.

\section{The Green's function and renormalization group calculations}\label{app_rg}

For the abrupt jump of Sec.~\ref{sec_abrupt} the Green's function, at $\lambda=0$, can be calculated exactly.
We have
\begin{eqnarray}\label{ms}
&&G(x,y;\tau)=T\sum_m\e^{\im \omega_m\tau}G_m(x,y)\textrm{ and}\\
&&\bigg[\frac{\omega_m^2}{2g_xu_x}-\frac{\partial}{\partial x}\bigg(\frac{u_x}{2g_x}\frac{\partial}{\partial x}\bigg)\bigg]G_m(x,y)=\delta(x-y)\,.\nonumber
\end{eqnarray}
Solving this differential equation subject to the appropriate boundary conditions \cite{Maslov1995,Sedlmayr2012a} gives
\begin{eqnarray}\label{green}
G(x,y;\tau)&=&\langle\phi(x,0)\phi(y,\tau)\rangle\\ \nonumber&=&
-\frac{\bg}{\pi}\ln\left|\sinh\left[\pi T\left(\frac{|x|}{u_x}+\frac{|y|}{u_y}-i\tau\right)\right]\right|
\\ \nonumber&&+\frac{\mathcal{L}[x,y]g_x}{\pi}\ln\left|\frac{\sinh\left[\pi T\left(\frac{|x|}{u_x}+\frac{|y|}{u_y}-i\tau\right)\right]}{
\sinh\left[\pi T\left(\frac{|x-y|}{u_x}-i\tau\right)\right]}\right|\,.\\ \nonumber&&
\end{eqnarray}
We have introduced $2\mathcal{L}[x,y]\equiv 1+\sgn[x]\sgn[y]$ .

The renormalization procedure is done in the standard manner by expanding the perturbation,
 $\exp(-\int\ud\tau H')$, to first order
and integrating out  the fields with fast Fourier components near the band-edge $\Lambda'<|k|<\Lambda$.  In order  to recover the original
form after re-exponentiating the action we rescale $\Lambda\tau_{\textrm{new}}=\Lambda'\tau$,
and define the new coupling constant $\lambda$, as
\begin{equation}
\lambda(\Lambda')=\frac{\Lambda}{\Lambda'}\lambda(\Lambda)e^{-\pi \tilde G_>(x=y=\tau=0)}\,.
\end{equation}
where $\tilde G_>$ is the Green's function after integrating
out the fast modes.

Therefore for the RG equation what we need is the Green's function summed over the fast modes. First let us change variables to $r=x-y$ and $R=(x+y)/2$. Then, with $u(r,R=0)=2u_xu_{y}/(u_x+u_{y})|_{x=-y}=2u_\ell u_r/(u_\ell +u_r)\equiv u$, we have
\begin{equation}\label{tgr}
G(r,R=0;\tau)=-\frac{\bg}{\pi}\ln\left|\sinh\left[\pi T\left(\frac{|r|}{u}-i\tau\right)\right]\right|\,.
\end{equation}
This is the same as the Green's function for a homogeneous case, but with a new velocity and a new Luttinger parameter:
\begin{equation}
 \frac{1}{\bg}=\frac{1}{2}\left[\frac{1}{g_\ell}+\frac{1}{g_r}\right]\,.
\end{equation}
Now we require
\begin{eqnarray}
G_>(0,0;0)&=&\sum_{\Lambda'<|k|<\Lambda}G(k,R=\tau=0)\\\nonumber
&=&\sum_{\Lambda'<|k|<\Lambda}\int\ud re^{\im kr}G(r,R=\tau=0)\,.
\end{eqnarray}
Thus integrating out the fast Fourier components
results in a change of the
Green's function,
\begin{eqnarray}
 G_>\approx
\frac{\bg}{\pi} d\ln{\Lambda}\,,
\end{eqnarray}
which governs the renormalization in the usual manner:
\begin{equation}
\label{RG}
\frac{1}{\lambda}\frac{d\lambda}{d\ln{\Lambda}}=1-{\bg}\,.
\end{equation}
We therefore expect that the effective backscattering renormalizes as
a power law in the temperature $R\propto T^{\bg-1}$, which in turn
affects the conductance and other physical observables accordingly. This has been confirmed numerically.\cite{Sedlmayr2012a}

\section{Useful sums for the boundary terms}\label{app_sum}

To find the coefficients of the backscattering terms several sums are needed.  We want
\begin{equation}
I\equiv\sum_{\substack{x=ja\\j\in\mathbb{Z}}}\e^{-2\im k_{F,x}x}F(x)O(x)
\end{equation}
in the particular case where we can write
\begin{equation}
F(x)=\underbrace{F_\ell\hside(-x-a)}_{\equiv F_\ell(x)}+\underbrace{F_r\hside(x)}_{\equiv F_r(x)}
\end{equation}
with $\hside(0)\equiv 1$. Assuming $O_x$ is slowly varying on a length scale of $a$, this allows us to write
\begin{eqnarray}
I&\approx&O(x=0)\sum_{\substack{x=ja\\j\in\mathbb{Z}}}\sum_{i=1,2}\e^{-2\im k_{Fi}x}F_i(x)\\\nonumber
&\approx&O(x=0)\sum_{\substack{x=ja\\j\in\mathbb{Z}}}\sum_{i=1,2}\e^{-2\im k_{Fi}x}\frac{\mathcal{Z}_i}{2}[F_i(x)-F_i(x+a)]\,,
\end{eqnarray}
with $\mathcal{Z}_i=1+i\cot[k_{Fi}a]$. Only a single term of each sum over $x$ is non-zero and we find
\begin{equation}
I\approx O(x=0)\left[\frac{\im F_\ell e^{\im k_{F\ell}a}}{2\sin[k_{F\ell}a]}-\frac{\im F_re^{\im k_{Fr}a}}{2\sin[k_{Fr}a]}\right]\,.
\end{equation}

We may also be interested in the case where
\begin{equation}
F(x)=\underbrace{F_\ell\hside(-x-a)}_{\equiv F_\ell(x)}+\underbrace{F_r\hside(x-a)}_{\equiv F_r(x)}+F_0\delta(x)
\end{equation}
and we can independently change $F_0$ on the central site. Then with $I\approx O_{x=0}\lambda$ we find
\begin{equation}
\lambda=F_0+\frac{\im F_\ell e^{\im k_{F\ell}a}}{2\sin[k_{F\ell}a]}-\frac{\im F_re^{-\im k_{Fr}a}}{2\sin[k_{Fr}a]}\,.
\end{equation}

\section{The mode expansion and its correlation functions}\label{app_mode}

Let's first consider the solutions of the classical equation of
motion, Eq.~\eqref{classicalem}, subject to the boundary conditions
Eq.~\eqref{bc1}.  There are two types of oscillating solutions
\begin{equation}
\phi_k^{(1)}(x,t)\sim\e^{\im kt}\cos(kx) \label{cos}
\end{equation}
with $\partial_x\phi(x=0)=0$,
and
\begin{equation}
\phi_k^{(2)}\sim g_x\e^{\im kt}\sin(kx)\label{sign}
\end{equation}
with $\phi(x=0)=0$.  In addition there are solutions linear in $x$ and
$t$. The solutions linear in $x$ have the form
\begin{equation}
\phi^x(x,t)\sim g_x x\,.
\end{equation}
The bosonization formula \eqref{bosform} implies that the bosonic
field $\phi$ is periodic with period $\phi+\sqrt{\pi}$. Furthermore we
are considering solutions on a ring with circumference $2L$ thus
$\phi(x)=\phi(x+2L)+\sqrt{\pi}n$. The oscillatory solutions of both
types therefore must have $k=\pi l/L$ for $l=1,2,3,\ldots$. For the
solutions linear in $x$ the same periodicity conditions imply
\begin{eqnarray}
\phi^x(x,t)&=&\frac{\sqrt{\pi}n}{2\tg L}xg_{x}\,.\label{wind}
\end{eqnarray}
Let's now consider the mode expansion.$\hat \Pi$ is canonically conjugate to $\phi_0$ and the normalization of each term is
fixed by requiring the canonical commutation relations to hold.For $g_\ell\neq g_r$, we may expand in solutions of the classical equations of motion, while
respecting the canonical commutation relations. This leads to the mode expansion given in Eq.~\eqref{mode2}.

Using the mode expansion we can first calculate the bosonic commutators in the ground state.
We find that
\begin{widetext}
\begin{eqnarray}
\textrm{Re}\,\langle\tphi(x,t)\tphi(x,0)\rangle
&=&-\frac{1}{\tg}\frac{1}{8\pi}\ln\left|2^4\sin[\pi (t-2x)/2L]\sin[\pi (t+2x)/2L]\sin^2[\pi t/2L]\right|\\\nonumber&&+\frac{\bg}{g_x^2}\frac{1}{8\pi}\ln\left|\frac{\sin[\pi (t-2x)/2L]\sin[\pi (t+2x)/2L]}{\sin^2[\pi t/2L]}\right|
\end{eqnarray}
and
\begin{eqnarray}
\textrm{Re}\langle\phi(x,t)\phi(x,0)\rangle
&=&-\frac{\bg}{8\pi}\ln\left|2^4\sin[\pi (t-2x)/2L]\sin[\pi (t+2x)/2L]\sin^2[\pi t/2L]\right|\\\nonumber
&&+\frac{g^2_x}{\tg}\frac{1}{8\pi}\ln\left|\frac{\sin[\pi (t-2x)/2L]\sin[\pi (t+2x)/2L]}{\sin^2[\pi t/2L]}\right|\,.
\end{eqnarray}
Finally we will also need
\begin{equation}
\textrm{Re}\,\langle\tphi(x,t)\phi(x,0)\rangle
=\frac{1}{8\pi}\left[\frac{g_x}{\tg}+\frac{\bg}{g_x}\right]\ln\left|\frac{\sin[\pi (t-2x)/2L]}{\sin[\pi (t+2x)/2L]}\right|\,.
\end{equation}
\end{widetext}

\end{document}